\documentclass[10pt,journal,compsoc]{IEEEtran}

\ifCLASSOPTIONcompsoc
  \usepackage[nocompress]{cite}
\else
  \usepackage{cite}
\fi
\usepackage{cutwin}
\ifCLASSINFOpdf
\else
\fi

\usepackage{multirow}
\usepackage{subfigure}

\usepackage{epsfig}
\usepackage[ruled,vlined]{algorithm2e}
\newcommand\MYhyperrefoptions{bookmarks=true,bookmarksnumbered=true,
pdfpagemode={UseOutlines},plainpages=false,pdfpagelabels=true,
colorlinks=true,linkcolor={black},citecolor={black},urlcolor={black},
pdftitle={Bare Demo of IEEEtran.cls for Biometrics Council Journals},
pdfsubject={Typesetting},
pdfauthor={Michael D. Shell},
pdfkeywords={Biometrics Council, IEEEtran, journal, LaTeX, paper,
             template}}
\ifCLASSINFOpdf
\usepackage[\MYhyperrefoptions,pdftex]{hyperref}
\else
\usepackage[\MYhyperrefoptions,breaklinks=true,dvips]{hyperref}
\usepackage{breakurl}
\fi
\usepackage[none]{hyphenat}

\usepackage[dvipsnames]{xcolor}

\begin{document}
%

\title{GANTouch: An Attack-Resilient Framework for Touch-based Continuous Authentication System}
%
%
%
%

\author{Mohit Agrawal, Pragyan Mehrotra, Rajesh Kumar, and Rajiv Ratn Shah
\IEEEcompsocitemizethanks{\IEEEcompsocthanksitem Mohit Agrawal (mohita@iiitd.ac.in), Pragyan Mehrotra (pragyan18168@iiitd.ac.in), and Rajiv Ratn Shah (rajivratn@iiitd.ac.in) are with IIIT Delhi, India. Rajesh Kumar (rajesh.kumar@bucknell.edu) is with Bucknell University, USA. }}

%
%

\markboth{To appear in IEEE Transactions on Biometrics, Identity and Behavior (T-BIOM)}%
{IEEE Biometrics Council Journals}
%



\IEEEtitleabstractindextext{%
\begin{abstract}
Previous studies have shown that commonly studied (vanilla) implementations of touch-based continuous authentication systems (V-TCAS) are susceptible to active adversarial attempts. This study presents a novel Generative Adversarial Network assisted TCAS (G-TCAS) framework and compares it to the V-TCAS under three active adversarial environments viz. Zero-effort, Population, and Random-vector. The Zero-effort environment was implemented in two variations viz. Zero-effort (same-dataset) and Zero-effort (cross-dataset). The first involved a Zero-effort attack from the same dataset, while the second used three different datasets. G-TCAS showed more resilience than V-TCAS under the Population and Random-vector, the more damaging adversarial scenarios than the Zero-effort. On average, the increase in the false accept rates (FARs) for V-TCAS was much higher (27.5\% and 21.5\%) than for G-TCAS (14\% and 12.5\%) for Population and Random-vector attacks, respectively. Moreover, we performed a fairness analysis of TCAS for different genders and found TCAS to be fair across genders. The findings suggest that we should evaluate TCAS under active adversarial environments and affirm the usefulness of GANs in the TCAS pipeline. 
\end{abstract}
\begin{IEEEkeywords}
Continuous Authentication, Behavioral Biometrics, Touchstrokes, Adversarial Attacks, Fairness, and GANs. \\ 

©2022 IEEE. Personal use of this material is permitted. Permission from IEEE must be obtained for all other uses, in any current or future media, including reprinting/republishing this material for advertising or promotional purposes, creating new collective works, for resale or redistribution to servers or lists, or reuse of any copyrighted component of this work in other works.
\end{IEEEkeywords}}
\maketitle

\IEEEdisplaynontitleabstractindextext


%
\IEEEpeerreviewmaketitle

\section{Introduction}
\label{sec:introduction}
\IEEEPARstart{U}ser authentication is an established area of research. Various methods have been used, including PIN, password, fingerprint, and face. With the changing landscape of human-computer interaction, the need for non-intrusive and continuous authentication systems is rising and evident. The study of behavioral footprints originated from human-computer-interaction for identity authentication has become an interesting area of research over the past decade. Among several behavioral footprints (e.g., gait \cite{OneClassRajesh2018}, keystroke \cite{kumar2016continuous}, touchstroke \cite{kumar2016continuous}, voice \cite{VoiceContAuth}, body-movements \cite{IJCB2017}), touch-strokes have been widely studied and have shown promise for non-intrusive continuous authentication \cite{frank2012touchalytics, TouchFirstAuth, serwadda2013verifiers, ModalSwipeContinuous, patel2016continuous,sitova2015hmog,kumar2016continuous}.

\subsection{Continuous Authentication via TCAS} One of the significant drawbacks of traditional means (e.g., fingerprint, face, PIN, password) is that they require user attention. In other words, they are intrusive. The other significant weaknesses are that they offer only entry point authentication. One can use coercion, intoxication, social engineering, or other means to unlock the device and use it afterward. Thus, the researchers have been focusing on developing non-intrusive continuous authentication systems that are user-friendly and \textit{resilient to active adversaries}. The non-intrusiveness here means that the authentication system would require minimal or no attention from users to verify their identity. The continuous part implies that the user's identity would be verified at frequent time intervals, either fixed or triggered by user actions on the device. Multiple continuous impostor alarms would lock users out of the system or require users to present additional credentials to continue. The frequency of the authentication and the number of impostor alarms required to lock users out of the system (say $n_a$) depend on the application scenario. For example, the frequency of the authentication would be relatively high in a high-security environment (e.g., military bases), and $n_a$ would be low. In contrast, in a low-security environment (e.g., a defensive driving course), the frequency of authentication would be kept low and $n_a$ high. 

The continuous authentication error rate need not be close to zero (i.e., comparable to fingerprint or face), which is often expected from an entry-point authentication system because an attacker will have to bypass multiple checks during a meaningful adversarial session of a continuous authentication system. The chances of the attacker doing so would be $p*q^n$ for n number of checks assuming $q$ is the probability of bypassing a check in a continuous authentication setup, and initial verification was using the entry-point authentication system. The value of $p*q^n$ will decline exponentially and catch up or will become lower than the probability of the attacker fooling the entry-point authentication system \cite{lee2017method,lee2017methodpaper}. Besides, continuous authentication systems do not have to replace entry-point authentication systems. Instead, they can be deployed with entry-point authentication systems to achieve higher security and usability.

The suitability of touchstrokes for non-intrusive continuous authentication has been credited to its universality, collectability, distinctiveness, acceptability, permanence, performance, and difficulty in reproduction by someone else \cite{frank2012touchalytics, 2018Benchmark, BehavioralBiometricsAcceptability, ForgeryResistantTCASFrank}. The authentication systems proposed in most previous studies consisted of a typical machine learning pipeline, i.e., data collection, feature engineering, classification, and performance evaluation. The introductory studies \cite{frank2012touchalytics, TouchFirstAuth, Li2013UnobservableRF} focused primarily on collecting touchstrokes while users browsed pages, images, and answered questions; extracting a set of features from each touch stroke; using the feature vectors to train and test authentication models, and evaluating the models using genuine reject, i.e., False Reject Rate (FRR) and impostor accept, i.e., False Accept Rates (FAR). Later studies \cite{serwadda2013verifiers,ModalSwipeContinuous,kumar2016continuous,sitova2015hmog,2018Benchmark,RamaChellappa} explored several variants such as usage contexts (sitting and walking) \cite{sitova2015hmog}, separate templates for different types (left, right, up, and down) of swipes \cite{serwadda2013verifiers}, fusion with phone movements \cite{kumar2016continuous,OneClassRajesh2018}, and bench-marking different classifiers on multiple datasets \cite{2018Benchmark}. 

The majority of the previous studies have treated identity authentication as a two-class problem \cite{frank2012touchalytics,serwadda2013verifiers,2018Benchmark}. In contrast, the rest have considered it a one-class classification problem \cite{OneClassRajesh2018}. Two-class classification-based approached achieved lower authentication error rates \cite{OneClassRajesh2018}. The performance of authentication systems was reported in terms of Equal Error Rate (EER--a point on ROC where FAR and FRR are equal) \cite{frank2012touchalytics, sitova2015hmog} or \cite{OneClassRajesh2018} Half Total Error Rates (HTER--an average of false accept and false reject rates) \cite{bengio2002confidence}. EER is generally used for setting the threshold during training/validation as we cannot change the threshold during testing. HTER is recommended for reporting testing performance \cite{bengio2002confidence,poh2006HTER}. To summarize, previous studies suggested that touchstrokes are a viable means for non-intrusive continuous authentication and have reported average error rates around $10\%$ percent, which could be good enough, especially for continuous authentication in the civilian domain \cite{chellappa2019continuous}. The problem, however, is that the studies have assumed the non-existence of active adversaries. Since the data generated or stored on smart devices are invaluable, active adversaries would likely exist. The following section describes some adversarial scenarios. 

\subsection{Adversarial Scenarios for TCAS}
The plausible attacks on TCAS can be grouped into three categories based on time, expertise, and equipment that the attackers expend \cite{EffortAsAFactor,frank2012touchalytics,serwadda2013verifiers,kumar2016continuous, RandomAttackMutibiometric, MimicryAttackOnSwipes, RoboticRobbery}. In the specific case of TCAS, an attacker needs to meet one or more of the following requirements. \textbf{R1:} ability to inject data into the authentication pipeline \cite{RandomAttackMutibiometric}, \textbf{R2:} access to the target's biometric samples \cite{RoboticRobbery}, and \textbf{R3:}  reproduction of samples by training imitators (human, machine, or human+machine) in real-time \cite{MimicryAttackOnSwipes}. Based on the amount of effort needed to meet these requirements, ongoing discussion on the Strength of Function for Authenticators by the National Institute of Standards and Technology (NIST) \cite{EffortAsAFactor}, and a recent survey \cite{AttackSurvey2021}, we group the adversarial scenarios into the following three groups.

\subsubsection{Minimal-effort attack} Attackers need to meet the first requirement i.e. \textbf{R1:} ability to inject data into the authentication pipeline \cite{RandomAttackMutibiometric,frank2012touchalytics,serwadda2013verifiers,kumar2016continuous,RoboticRobbery}, to launch attacks falling under this category. Attack strategies that have been studied for TCAS in the past and fall under this category are listed and described below: 

\textit{Zero-effort attack:} under this attack, the attack vectors are randomly borrowed from each possible impostor (users other than the genuine user). The name zero-effort means that the impostors made zero effort to copy or imitate the genuine user. Consequently, zero-effort is one of the most widely adopted and studied attacks, primarily due to convenience. Consequently, most previous studies have evaluated and reported TCAS's performance under zero-effort attack \cite{serwadda2013verifiers,frank2012touchalytics,kumar2016continuous}. Therefore, the performance under a zero-effort attack is considered the baseline performance of TCAS in this paper.

\textit{Population-based attack:} in this attack, the attack vectors are generated by taking feature vectors from all possible impostors into account. For example, attack vectors can be created by taking means of each of the individual features \cite{RoboticRobbery}. It is similar to creating a master key (like creating a master face by averaging all possible faces available to breach face-based authentication). Alternatively, one can create multiple groups of impostors using clustering techniques to create multiple master keys as suggested in \cite{populationattackGait} for gait-based biometrics. This type of attack assumes that the attackers have access to public datasets. 

\textit{Random-vector attack:} 
in this attack setup, the attackers generate random attack vectors by utilizing prior knowledge, i.e., the length of the feature vectors and range of feature values. Researchers often scale feature values in a fixed range, such as k-nearest neighbors demand so, as some classification algorithms. Zhao et al. \cite{RandomAttackMutibiometric} evaluated the impact of a Random-vector attack on TCAS and concluded that a Random-vector attack is highly effective on behavior-based authentication systems. 
 
\subsubsection{Moderate-effort attack} This category of attacks is required to meet the first two criteria, i.e., \textbf{R1:} ability to inject data into the authentication pipeline and \textbf{R2:} access to the target's biometric samples. The performance of TCAS against this category of attacks has not been reported. This attack category has been mostly studied on behavioral biometrics other than touchstrokes. We include the description of this attack category for completeness.

\textit{Snoop-forge-replay:} This kind of attack has been studied in the context of keystroke-based continuous authentication systems \cite{Snoop-forge-reply-keystroke}. As the name suggests, this attack works in three steps. \textit{Snoop:} the attacker uses social engineering or other possible means to gain access to the biometric samples (or feature vectors). \textit{Forge:} the attacker reproduces desired number of attack vectors by using the stolen genuine samples. \textit{Replay:} the attacker then replays/feeds the forged samples to the authentication API for the time it wants to gain access to the resources protected by the continuous authentication system. 

\subsubsection{High-effort attack} This category of attacks is difficult to launch as the attackers need to meet \textbf{R2}: access to the target's biometric samples and \textbf{R3}: reproduction of samples by training imitators in real-time. These attacks are difficult to detect because they do not require any modifications to the device or leave any footprints to be traced later. These attacks involve training individuals or machines (robots) or one with the help of the other to reproduce samples that are close enough to that of the target or an average gestures derived from publicly available databases \cite{MimicryAttackOnSwipes, RoboticRobbery}. Serwadda et al. \cite{RoboticRobbery} trained a Lego robot to match the swipes of the target user. In comparison, Khan et al. \cite{MimicryAttackOnSwipes} trained human imitators to produce the attack vectors. One could combine and attack TCAS using a robot imitator assisted by humans or robots. 

\subsection{Possible Countermeasures}
Previous studies such as \cite{DistanceBasedMatchersAreImmuneToRandomAttacks} have suggested that biometric systems based on raw data level distance-based matching are more resilient to the attacks; however, they exhibit very high error rates; in general, \cite{RandomAttackMutibiometric,rajesh2020, MyPhDThesis}. On the other hand, machine learning-based matches achieve much lower error rates; therefore, they are heavily used for implementing TCAS \cite{2018Benchmark, LatestSurveyTouchUsabilitySecurity2020, serwadda2013verifiers} than the distance-based matches. The primary issue with the previous machine learning-based TCAS implementation is that they assumed that Random-vector would belong to the impostor class. However, Zhao et al. \cite{RandomAttackMutibiometric} argued that Random-vectors might also belong to the genuine class. Therefore, we chose to evaluate the machine learning-based implementations of TCAS under the most common adversarial scenarios. 

Two approaches have been suggested as a possible defense against minimal effort attacks on machine learning-based TCAS. The first utilizes two Generative Adversarial Networks (GANs) \cite{goodfellow2014generative, IJCB2021} (see Figure \ref{AdversarialTouch}). The second focuses on reducing the acceptance region by generating synthetic data (noise) around the genuine samples and considering the generated data as impostors \cite{RandomAttackMutibiometric}. The reduced acceptance region decreases the chances of classifying any Random-vector as genuine.  

To summarize, there are multiple ways to implement and attack TCAS. This paper focuses on machine-learning-based implementations of TCAS and its evaluation under minimal-effort attack scenarios, including zero-effort, Population, and Random-vector attacks on multiple datasets under the same and cross dataset scenarios. We could launch and evaluate only minimal-effort attacks in this paper. The evaluation of the moderate or high-effort attack is a tedious task as it requires attack data collection from trained human or robot imitators. We plan to investigate this part in the future.

\subsection{Main Contributions}
The main contributions are as follows: 
\begin{itemize}
    \item We summarize traditional TCAS implementations, establish a classification of possible adversarial scenarios for TCAS, and summarize the existing defense approaches.
    
    \item We implement a traditional machine learning pipeline for TCAS. We refer to the same as vanilla TCAS (V-TCAS) hereafter and test the same under three adversarial scenarios, viz. Zero-effort, Population-based, and Random-vector. V-TCAS's false acceptance increased significantly under these attack scenarios.
    
    \item Next, we implement (extend the idea presented in the conference paper \cite{IJCB2021}) a novel Generative Adversarial Networks assisted TCAS framework (G-TCAS) and test the same under the three aforementioned adversarial scenarios. The results suggest that G-TCAS is more resilient to adversarial environments than V-TCAS.
    
    \item We benchmarked four widely studied classifiers (each with a diverse learning paradigm). The superiority of G-TCAS over V-TCAS was evident across the experimental setups. 
    
    \item Additionally, we analyze the fairness of V-TCAS and G-TCAS using kernel density plots, only to find out that TCAS is fair among different genders. 
\end{itemize}

The rest of the paper is structured as follows. Section \ref{RelatedWorks} discusses the closely related works. Section \ref{DesignOfExperiments} presents the design of experiments. Section \ref{ResultsAndDiscussion} presents and discusses the results, respectively. Finally, we conclude the paper and provide future research directions in Section \ref{secConclusionAndFutureWork}. The code is publicly available\footnote{\href{https://github.com/midas-research/GANTouch-TBIOM}{https://github.com/midas-research/GANTouch-TBIOM}}.

\section{Related work}
\label{RelatedWorks}
This work extends the idea presented in the conference paper \cite{IJCB2021} in the following dimensions: 

\textit{Additional attack scenario:} The conference paper evaluated V-TCAS and G-TCAS under Zero-effort and Population-based adversarial scenarios. We extend the analysis to random-input attacks, which were shown to be very effective in penetrating V-TCAS in a recent study \cite{RandomAttackMutibiometric}. 

\textit{Additional datasets:} In the conference paper \cite{IJCB2021}, we had used only two datasets viz. Serwadda-touch and BBMAS-Touch. In this paper, we include two more datasets, viz. Hand Movements, Orientation, and Grasp (HMOG), and UMDAA-02Touch Datasets. The additional datasets were used to create the population-based attack environment.  

\textit{Gender-wise fairness analysis and statistical significance:} To the best of our knowledge, no prior study has explored whether TCAS discriminates between different genders. Therefore, we conduct additional analysis on V-TCAS and G-TCAS to determine whether TCAS is fair across gender. Unfortunately, the analysis consisted of males and females and the BBMAS-Touch dataset. Because none of the publicly available touch stroke datasets included contained gender information (to the best of our knowledge). We also include a series of kernel density plots to show that the results and conclusion hold across users and gender groups. 

Besides \cite{IJCB2021} and the literature cited in it, we review papers that have investigated the vulnerability of TCAS. Additionally, we discuss studies that have utilized GANs as a countermeasure beyond TCAS. 

Zhao et al. \cite{RandomAttackMutibiometric} demonstrated that previously studied designs of behavioral pattern-based authentication systems, including TCAS, are susceptible even to uniform Random-vectors. Their investigation showed that the acceptance region of the machine learning models is much bigger than the one occupied by the genuine samples. In other words, the probability of the Random-vector being accepted as a genuine is much higher than the false accept rate of the model. They demonstrated that if attackers know the length of the feature space and the range of values each feature takes, they can launch a successful attack by generating random feature vectors. The idea was evaluated on gait, touch, face, and speech. Results showed that the random attack was very successful in the case of TCAS. This paper also motivated us to evaluate our defense scheme against random attacks. 

Deb et al. \cite{GANTouch2020} conducted a preliminary study in which they applied Auxiliary Classifier Generative Adversarial Network (AC-GAN) to achieve error rates between 2 to 11\% under synthetic data attack. The study, however, included only ten users randomly selected from the Touchalytics dataset, consisting of 41 users. Nonetheless, the paper suggested that there is a promise in using GANs to train user authentication models. Gomez-Alanis et al. \cite{GANBA2022} concluded that GAN-based automatic speaker verification models are more robust against original and adversarial spoofing attacks. Last but not least, GAN-based defense has been applied in the image classification area to enhance the robustness of classification models against black-box and white-box adversarial attacks \cite{DefenseGAN2018}. 

\begin{table}[htp]. 
\small
\centering
\caption{The list of datasets used in this study, \# of subjects, avg. number of (valid) swipe gestures per user, and demographics availability.}
\vspace{0.03in}
\label{tab:samplestat}
\begin{tabular}{c|c|c|c}
\hline
\multicolumn{1}{c|}{\textbf{Dataset}} & \multicolumn{1}{c|}{\textbf{Users}} & \multicolumn{1}{c|}{\textbf{Swipe/User}} & \multicolumn{1}{c}{\textbf{Demographics?}} \\ 
\hline
BBMAS-Touch       &        117     &  173  &        Yes\\
\hline
Serwadda & 190 & 273 & No \\
\hline
HMOG & 100 & 2275 & No \\
\hline
UMDAA-02Touch & 48 & 4063 & No
              \\ 
\hline
\end{tabular}
\end{table}

\section{Design of Experiments}
\label{DesignOfExperiments}
Most of the previously proposed implementations of TCAS consist of a typical machine learning pipeline, including data collection and preprocessing, feature engineering, classification, and performance evaluation. The details of these steps are as follows.
\subsection{Datasets}
This study primarily uses BBMAS-Touch \cite{belman2019insights}. The decision to use this dataset as the main dataset was based on several factors, such as it has a good number of users ($117$), the number of swipes per user ($173$), and gender information which was desired for gender/fairness analysis, and the data were collected in a realistic setup. Three more datasets, viz. Serwadda \cite{serwadda2013verifiers}, HMOG \cite{sitova2015hmog}, and UMDAA-02Touch \cite{RamaChellappa} were used to create cross dataset zero-effort adversarial environment. A summary of these datasets is provided in Table \ref{tab:samplestat} besides briefly describing each dataset in the following paragraphs.

\textit{BBMAS-Touch\cite{belman2019insights}:} This dataset consists of the touch portion of Syracuse University and Assured Information Security-Behavioral Biometrics Multi-Device, and Multi-Activity Data (SU-AIS BB-MAS) \cite{belman2019insights}. The participants were handed over a phone loaded with the data collection app during the data collection. The participants typed two fixed sentences. 
 
Then the participants presented a series of ten questions that required varying cognitive loads to be answered with a minimum of 50 characters. The exercise required the participants to swipe between questions. The data collection app implicitly recorded touch, keystroke, and corresponding movements (accelerometer and gyroscope readings) throughout the process.  

\textit{Serwadda\cite{serwadda2013verifiers}:} This dataset contains touch gestures collected from $190$ participants. The participants used \textit{Google Nexus S}, an Android-based smartphone, to answer a series of multiple-choice questions after reading or scrolling through images and textual paragraphs. Browsing through the passages and images to answer the questions generated hundreds of touch gestures. The data collection exercise consisted of two independent sessions, separated by at least a day. Each participant generated $400$ swipes on average. 

\textit{HMOG \cite{sitova2015hmog}:} This dataset was created using $10$ \textit{Samsung Galaxy S4}, Android-based phones. A total of $90$ individuals participated who were randomly assigned a reading, writing, or map navigation session. Each session lasted about 5-15 minutes. The participants were either sitting or walking while working on each session. Every participant performed 24 sessions (eight for each reading, writing, and map navigation session). The recorded signals consisted of raw touch events, tap gestures, scale gestures, scroll gestures, fling gestures, keypresses, and corresponding device movements captured by inertial sensors, viz. accelerometer, gyroscope, and magnetometer. 

\textit{UMDAA-02Touch \cite{RamaChellappa}:} This dataset consists of swipe gestures collected from $48$ participants for two months. The participants were not given any particular task to generate swipes. This dataset, thus, consists of touch gestures closer to how users interact with the phone through touch. Google Nexus 5, an Android-based phone, was used in the data collection. The data was collected for over two months, unrestricted. 
\begin{table*}[htp]
\centering
\footnotesize
\caption{The list of all the features extracted from individual swipes.}
\begin{tabular}{|c|c|c|c|}
\hline
\multicolumn{1}{|c|}{\textbf{Feature Id}} 
& \multicolumn{1}{|c|}{\textbf{Feature name}}                                                        & \multicolumn{1}{c|}{\textbf{Equation}} 
                     & \multicolumn{1}{c|}{\textbf{Description}}
\\ \hline
$1$ & $swipe\_duration$ & $t_{end}$ - $t_{start}$   & Duration between start and end of a swipe \\ \hline

$2-5$ & \begin{tabular}[c]{@{}l@{}}$start\_x$, $start\_y$, \\ $end\_x$, $end\_y$ \end{tabular}                & $x_0$, $x_0$, $x_{n-1}, $y$_{n-1}$ & Coordinates of a swipe

\\ \hline
$6$ & $dp$                                                                                            & $\sqrt{(x_{n-1} - x_0)^2 + (y_{n-1} - y_0)^2}$ & Displacement of swipe \\ \hline

$7$ & $l$                                                                                           & l = $\sum_{i = 1}^{ n - 1} \sqrt{(x_{i-1} - x_i)^2 + (y_{i-1} - y_i)^2}$   & Length of the swipe                                                                         \\ \hline
$8$ & $velocity$                                                                                      & $dp$/($t_{end} - t_{start}$)    & Velocity of a swipe                                                          \\ \hline
$9$ & $initial\_v$                                                                                    & Velocity of first 5\% of the points & Initial velocity                                                \\ \hline
$10$ & $final\_v$                                                                                      & Velocity of final 5\% of the points & Final velocity                                                    \\ \hline
$11$ &  $mean\_v$                                                                                      & 
$(v_x)_i =  \frac{x_i - x_{i-1} }{t_i - t_{i-1}}, (v_y)_i =  \frac{y_i - y_{i-1} }{t_i - t_{i-1}}$   
 & Pairwise average velocity (magnitude)                                                       \\ \hline
$12$ &  $direction$                                                                                     & $\theta= tan^{-1} (\frac{x_{n-1} - x_0}{y_{n-1} - y_0})$ & Angle of line joining start and end points                                                \\ \hline
$13$ & $area$                                                                                        & $A = {1}/{n} \times \sum_{i=1}^{n}\pi \times a_i \times b_i$  & Average area of the fingertip over the swipe                                                \\ \hline
$14$ & $acceleration$                                                                                  & $(a_x)_i = { \frac{(v_x)_{i} - (v_x)_{i-1} }{t_i - t_{i-1}}}, (a_y)_i = { \frac{(v_y)_{i} - (v_y)_{i-1} }{t_i - t_{i-1}}}$ & Acceleration between start and end points                                                   \\ \hline
$15$ & $mean\_a$                                                                                      &  Pairwise average acceleration (magnitude) & Average acceleration                                                  \\ \hline
$16$ & $initial\_a$                                                                                    & Acceleration of first 5\% points & Initial acceleration                                                    \\ \hline
$17$ & $final\_a$                                                                                      & Acceleration of final 5\% points & Final acceleration                                                       \\ \hline
$18-20$ & $aP_{25}$, $aP_{50}$, $aP_{75}$                                                                              &  Acceleration ($aP_m$) of $m$\% swipe  & Acceleration percentile                                                \\ \hline
$21-23$ & $vP_{25}$, $vP_{50}$, $vP_{75}$                                                                              & Velocity ($vP_m$) of $m$\% swipe  & Velocity percentile                                                    \\ \hline
$24$ & $speed$                                                                                         & $l$/($t_{end} - t_{start}$)  & Speed of a swipe                                                                        \\ \hline
$25-26$ & $initial\_s$, $final\_s$                                                                         & Speed of first 5\% points, Speed of final 5\% points  & Initial speed, Final speed                                                                  \\ \hline
$27-29$ & $sP_{25}$, $sP_{50}$, $sP_{75}$                                                                              &  Speed ($sP_m$) of $m$\% swipe      & Speed percentile                                                   \\ \hline
$30-34$ & \begin{tabular}[c]{@{}l@{}}$mean\_v_x$, $mean\_v_y$, \\ $mean\_a_x$, $mean\_a_y$, \\ $mean\_d$\end{tabular} & Average of $v_x$, $v_y$, $a_x$, $a_y$, $dp$     & Mean of features                                                       \\ \hline
$35$ & $max\_d$                                                                                    & $d_i = \frac{|y_i - m \times x_i - c|}{\sqrt{1 + m^2}}$   & Maximum of deviations                                                                       \\ \hline
$36-38$ & $v_xP_{25}$, $v_xP_{50}$, $v_xP_{75}$                                                                           & \begin{tabular}[c]{@{}l@{}}Velocity ($v_xP_m$) of $m$\% swipe\end{tabular}  & Mean velocity percentile   \\ \hline
$39-41$ &  $v_yP_{25}$, $v_yP_{50}$, $v_yP_{75}$                                                                           & \begin{tabular}[c]{@{}l@{}} Veocity ($v_yP_m$) of $m$\% swipe\end{tabular}  & Mean velocity percentile   \\ \hline
$42-44$ & $a_xP_{25}$, $a_xP_{50}$, $a_xP_{75}$                                                                           & \begin{tabular}[c]{@{}l@{}} Acceleration ($a_xP_m$) of $m$\% swipe\end{tabular} & Mean acceleration percentile\\ \hline
$45-47$ & $a_yP_{25}$, $a_yP_{50}$, $a_yP_{75}$                                                                           & \begin{tabular}[c]{@{}l@{}} Acceleration ($a_yP_m$) of $m$\% swipe\end{tabular} & Mean acceleration percentile\\ \hline
\end{tabular}
\label{touchfeatures}
\end{table*}
 
\subsection{Separation of Training and Testing Data}
BBMAS-Touch, the base dataset, was divided into two parts with a 60:40 ratio, with 60\% being the training dataset while 40\% for testing. The training dataset was used for training the model and deciding upon the values of hyperparameters using a 5-fold cross-validation technique. The test data was kept unseen during the training phase. The testing environment used Serwadda, HMOG, and UMDAA-02Touch datasets. 

\subsection{Preprocessing and Feature Engineering}
We excluded the swipes that had five or fewer data touchpoints. This process resulted in swipes that likely consisted of unique individual behavior. For BBMAS-Touch, the preprocess step removed about 10.33\% of swipes resulting in 20286 swipe gestures. Other datasets were cleaned similarly. UMDAA-02Touch did not have the pressure information, so we appended zero. The next step was to extract features from the swipes to derive hidden characteristics of the swipe gestures. A swipe gesture $S$ can be defined as a set of tuples representing $n$ touch events between touching the screen with fingers and lifting the fingers from the screen ans represented as follows:
\begin{equation}\label{e0}
S = (x, y, t, a, b)_{\texttt{i=1 to n}}
\end{equation} where $x, y, t, a,$ and $b$ represent, x-coordinate, y-coordinate, time, and major-axis and minor-axis of the fingertip of each touch event, respectively. 

Building upon previous studies \cite{frank2012touchalytics,serwadda2013verifiers}, we extracted $30$ features and added $17$ new features. The process resulted in $47$ features as listed and described in Table \ref{touchfeatures}.

\subsection{Class Imbalance} Since we used the rest of the users as impostors, the number of genuine feature vectors turned out to be far less than the number of impostor feature vectors. To address such as class imbalance, we used the Adaptive Synthetic Over-sampling approach for imbalanced learning (ADASYN) \cite{he2008adasyn} after trying several variants of Synthetic Minority Oversampling Technique (SMOTE) \cite{chawla2002smote}. 

\subsection{Choice of Classifiers} Previous studies have used several classifiers. For example, Serwadda et al. \cite{serwadda2013verifiers} evaluated ten classifiers. Frank et al. \cite{frank2012touchalytics} used SVM and k nearest neighbors, while Fierrez et al. \cite{2018Benchmark} used SVM, GMM, and their fusion. Kumar et al. \cite{kumar2016continuous,OneClassRajesh2018} experimented with eight classifiers. Previous studies did not agree on which classifier was the best, so we developed three criteria to select classifiers for evaluation. First, the classifiers should have achieved less
than 10\% error rates in the previous studies. Second, the classifier can be trained with small training data. The third criterion was the diversity of learning paradigms. The selection process resulted in Support Vector Machine (SVM), Random Forest (RF), and Multilayer Perceptron (MLP). We added Extreme Gradient Boosting (XGB) to the list primarily because it was not tested in the previous TCAS studies and has performed very well in online data science competitions. We used these many classifiers primarily because we wanted to ensure that the idea of including GAN in the training process is not limited to certain algorithms or learning paradigms. 

\subsection{The Continuous Framework}
The continuous part of the authentication systems was implemented using the sliding window scheme. A window initially contains $p$ consecutive swipes. The following windows are created by sliding the window that drops $q$ least recent swipes and adds $q$ following swipes. Preliminary experimentation led us to set $p$ and $q$ to $5$ and $1$, respectively. Instead of taking an average of the features as done in the past, we concatenated them, which resulted in $235$ features. These many features prompted us to use a mutual information-based feature selector in the pipeline. The number of features used for each user authentication model thus varied \cite{rajesh2020}. The varying number of features across the authentication offers another challenge for the attackers to figure out the length of the feature vectors in some adversarial environments \cite{RandomAttackMutibiometric}. In our implementations of those scenarios, we assumed that the attacker would have access to the feature vector length information.  

\begin{figure*}[htp]
    \centering
    \includegraphics[width=7.25in, height = 3.35in]{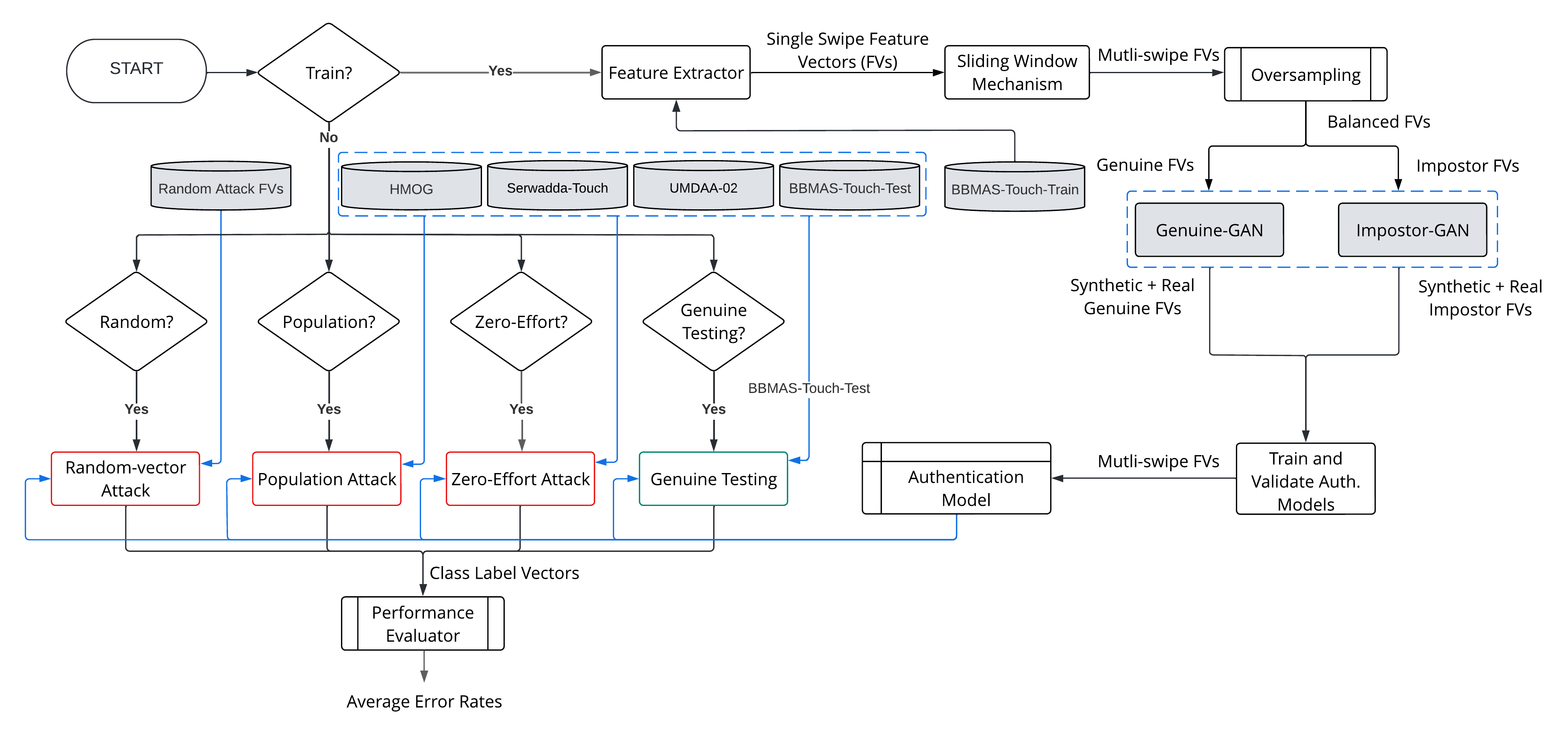}
    \caption{The system architecture of TCAS implemented in this paper. The novel part in the pipeline is highlighted with dotted blue lines, which makes use of two generative networks, viz., Genuine-GAN and Impostor-GAN. Additionally, we test the V-TCAS and G-TCAS under three adversarial environments viz., Zero-effort (traditional), Random-vector, and Population-based.}
    \label{AdversarialTouch}
\end{figure*}

\subsection{Training of V-TCAS and G-TCAS} The training of G-TCAS-based user authentication models is depicted in Figure \ref{AdversarialTouch}. In comparison, the training of V-TCAS-based authentication systems excludes the GAN-based component surrounded by blue dashed lines in the training pipeline. To train the authentication model $u_i$, we labeled the feature vectors extracted from the training session data of $u_i$ as genuine and the feature vectors extracted from the rest of the users, i.e., $U \setminus u_i$ impostors, where $U$ is the set of all the users. 

As rendered in Figure \ref{AdversarialTouch}, the G-TCAS framework uses a pair of Generative Adversarial Networks (GANs). The first viz. Genuine-GAN generates swipes similar (closer) to the Genuine swipes, while the second viz. Impostor-generated swipes are similar (closer) to the impostor swipes. Genuine-GAN consist of a discriminator $D(x_l)$ and a generator $G(z_l)$. Where $x_l$ represents feature vectors extracted from real swipes belonging to the genuine user. While $z_l$ represents the input noise for the generator. Both $D(x_l)$ and $G(z_l)$ are trained simultaneously as they play a min-max game with the value function given in Equation \ref{eq:eq9}.  
\begin{equation}
\label{eq:eq9}
\min_{G} \max_{D} V_l(D, G)= E_1 + E_2
\end{equation}
\begin{equation}
    E_1 = E_{{x} \sim p_{{l}}({x})}[\log D({x_l})]
\end{equation}
\begin{equation}
    E_2 = E_{{z_l} \sim p_{{z_l}}({z_l})}[\log (1-D(G({z_l})))]
\end{equation}
where, $p_l$ represents distribution of the generator over genuine swipes $x_l$, and $p_{zl}$ represent noise for generating (synthetic) genuine swipes. 

Similarly, Impostor-GAN consists of $G(z_a)$ and a discriminator $D(x_a)$. Where, $x_a$ represents feature vectors extracted from real swipes belonging to the impostors. While $z_a$ represents the input noise for the generator $G(z_a)$. Both $D(x_a)$ and $G(z_a)$ are trained simultaneously as they play a min-max game with the value function given in Equation \ref{eq:eq10}.   
\begin{equation}
\label{eq:eq10}
\min_{G} \max_{D} V_a(D, G)= E_3 + E_4
\end{equation}
\begin{equation}
E_3 = E_{{x} \sim p_{{a}}({x})}[\log D({x_a})]
\end{equation}

\begin{equation}
E_4 = E_{{z_a} \sim p_{{z_a}}({z_a})}[\log (1-D(G({z_a})))]
\end{equation}
where, let $p_a$ represents distribution of the generator over impostor swipes data $x_a$ and $p_{za}$ represent noise for generating (synthetic) impostor swipes. 

\begin{figure}[htp]
\centering
\begin{tabular}{cc}
\subfigure[Genuine (real vs. generated densities).]{\epsfig{file=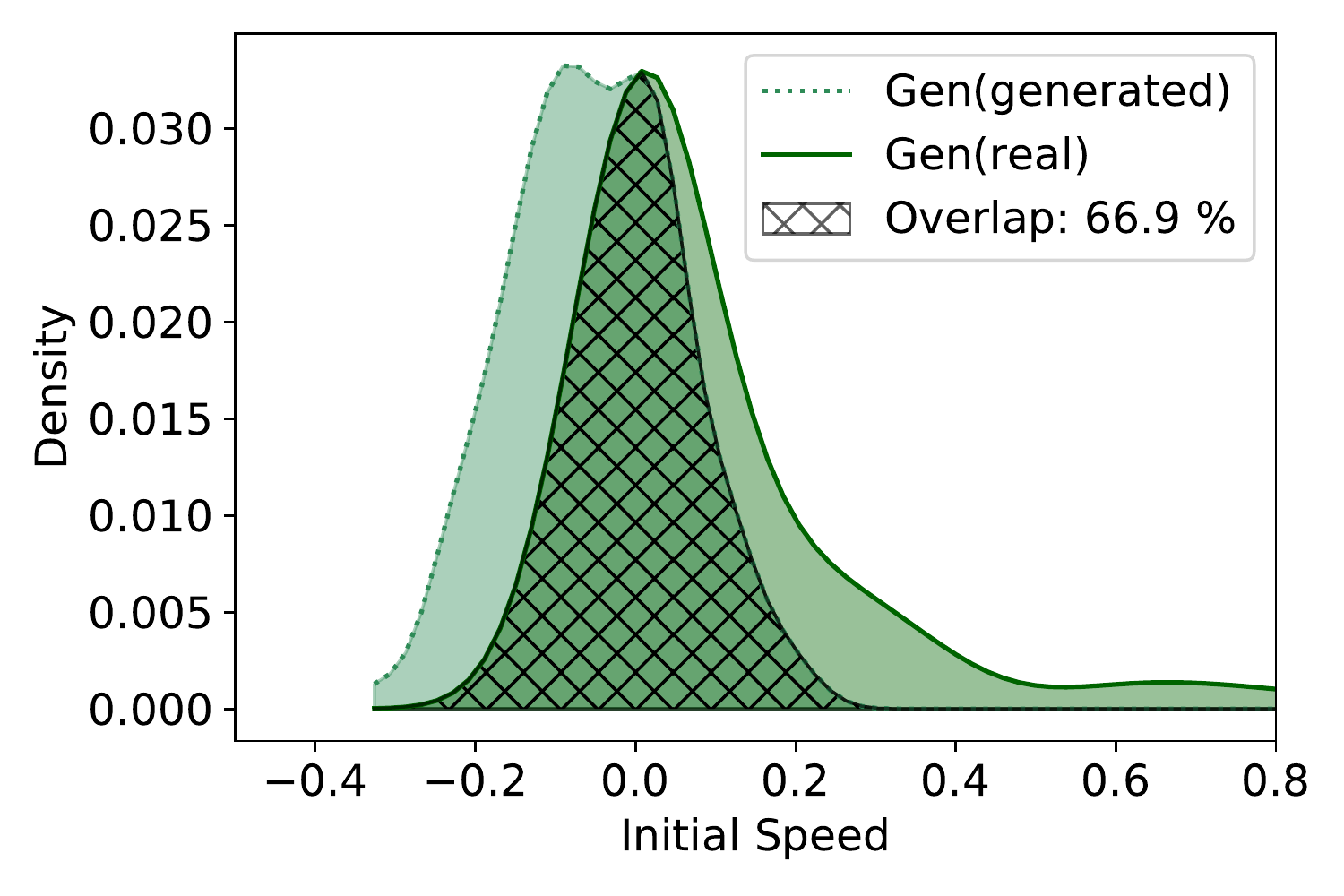, width=3in, height=1.75in}
\label{KDEPlotFinalSpeedGen14}} \\
\subfigure[Impostor (real vs. generated densities).]{\epsfig{file=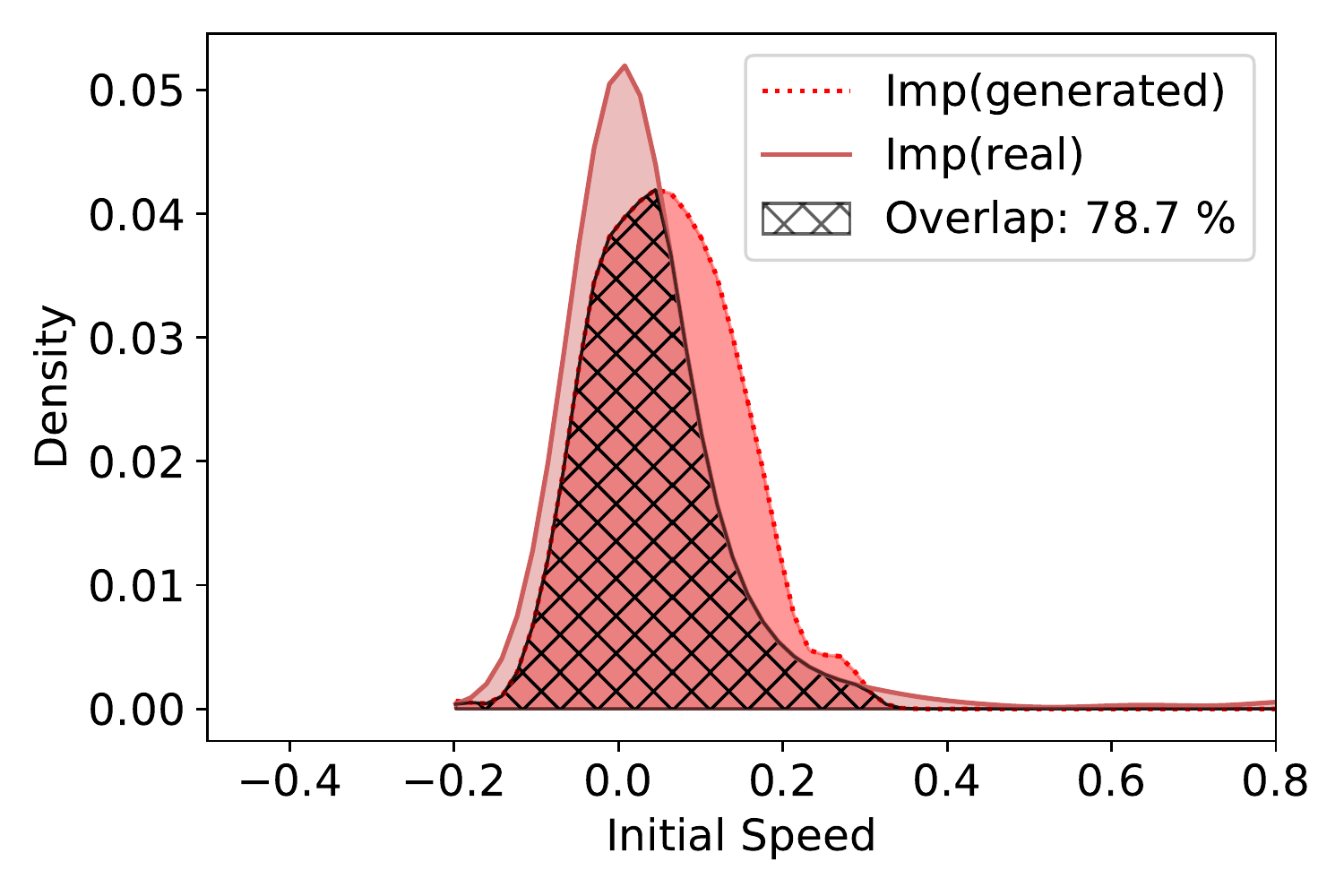, width=3in, height=1.75in}
\label{KDEPlotFinalSpeedImp}}
\end{tabular}
\caption{Visual evaluation of GAN generated data at the feature level. The illustration is only for demonstration purposes, and such overlap might not hold for all the features and users.} 
\label{GenuineGANKDEImpostorGANKDE}
\end{figure}

It is difficult to evaluate the quality of generated data by GANs, especially when we deal with non-visual data. Nevertheless, we examined the data generated by both Genuine-GAN and Impostor-GAN at the feature level using Kernel Density Estimation (KDE) plots. Figure \ref{GenuineGANKDEImpostorGANKDE} illustrates the distribution of real and generated feature values for one of the top features.


The outputs of both Genuine-GAN and Impostor-GAN are appended to genuine and impostors, respectively. Preliminary experiments suggested that the number of synthetic swipes included in the TCAS training/validation impacted the overall error rate, i.e., HTER. Therefore, we decided to consider the number of synthetic swipes as a hyperparameter during the model training and validation. The number of swipes that achieved the minimum HTER was selected. We experimented with (generated) swipes ranging between $[100, 1000]$ and found $250$ achieving the lowest validation HTER. 

\subsection{Testing of V-TCAS and G-TCAS} Once trained, each of the authentication models were tested for genuine fail and impostor pass. Test for genuine failure is straightforward. We test the models using the data (preferably collected in a different exercise than the training data) collected from genuine individuals. The percentage of failed genuine attempts is the Genuine or False Reject Rate (FRR). Test for the impostor pass, on the other hand, could be done in several possible ways. We test each model for impostor pass under three adversarial environments, viz. Zero-effort, Population, and Random-vector (see Figure \ref{AdversarialTouch}). The percentage of successful impostor attempts is called Impostor or False Accept Rates (FAR). We compute the FAR under each adversarial environment separately. The implementation of each adversarial environment is described below. 

\subsubsection{Zero-effort attack} This is the most widely adopted adversarial environment for TCAS. As the name suggests, the authentication models are tested against a dataset produced with no intention or effort to imitate or copy the genuine users. The traditional way to implement the zero-effort adversarial environment is to consider all users except the genuine user as impostors. Previous studies used users from the same dataset as impostors, so we refer to that scenario as the same dataset zero-effort environment. An impostor can come from anywhere, i.e., any dataset. So we implemented same-dataset as well as cross-dataset zero-effort attack scenarios. The step-by-step process is described in Algorithm \ref{algo_ze_attack}.

\subsubsection{Population attack} This adversarial environment was created in two steps. First, we computed the mean $(\mu_i)$ and the standard deviation $(\sigma_i)$ for each feature $(i)$ across all feature vectors from all the datasets except the dataset used in the training. Second, we generated feature vectors using the formula $\mu_i + r \times \sigma_i$ for each feature, where $r \in \mathcal{N}(0,3)$. We followed this process to generate $10000$ feature vectors to attack each authentication model. Algorithm \ref{algo_population_attack} explains the process in more detail.
\begin{algorithm} 
\small
\SetAlgoLined
\textbf{Input:} {A: list of authentication models for user $u_i \in  U$} \\
\quad    {D: List of datasets containing feature vectors} \\
\textbf{Output:} {C: List of dictionaries with keys (authentication models) and values (predicted labels)}\\
{$C \gets []$} \\
\For{dataset in D}{
{temp} $\gets$ \{\} \\
{X} $\gets$ {get\_feature\_matrix($u_i$, dataset)}\\
{X'} $\gets$ {normalize(X)} \\
\For{model in $A$}{
{pred\_labels $\gets$ []} \\
 \For{{feat\_vector in  X'}}{
 {pred\_labels.append(model.predict(feat\_vector))}}
 {temp[model] $\gets$ pred\_labels} \\
 }
 {C.append(temp)} \\
 }
 return C
\caption{{zero\_effort\_attack}(M[], D[], X[])}
\label{algo_ze_attack}
\end{algorithm}
\begin{algorithm} 
\small
\SetAlgoLined
\textbf{Input:} {A[]: list of authentication models for user $u_i \in  U$\\
\quad    M: list of means, computed over all the datasets\\
\quad    S: list of standard dev corresponding to the means}\\
\quad    N: number of feature vectors to be generated\\
\textbf{Output:} {C: A dictionary with keys (authentication models) and values (predicted labels)}\\
C $\gets \{ \}, {X'} \gets []$ \\

\For{$i \gets 0$ to $N$}{
 {attack\_vector} $\gets$ []\\
\For{{$\mu$, $\sigma$ in zip(M,S)}}{
    $r \gets \mathcal{N}(0, 3)$\\
    {attack\_vector.append}$(\mu[j] + r \times \sigma[j])$
 }
 {X'.append(attack\_vector)}
 }
{X' $\gets$ normalize(X')} \\
\For{model in $A$}{
{pred\_labels $\gets$ []} \\
 \For{{feat\_vector in  X'}}{
 {pred\_labels.append(model.predict(feat\_vector))}}
 {C[model]} $\gets$ {pred\_labels} \\
 }
  {return C}
\caption{{population\_attack}(A[], M, S, N)}
\label{algo_population_attack}
\end{algorithm}

\subsubsection{Random-vector attack} This adversarial environment was motivated from \cite{RandomAttackMutibiometric}. We selected uniform random values between $0$ and $1$ for each feature in the feature vector to implement this. We followed this process to generate $10000$ feature vectors to attack each authentication model. Algorithm \ref{algo_random_attack} shows the steps to implement the random-vector attack.

\subsection{Performance Evaluation}
To evaluate the performance of the authentication systems, we used FRR (defined in Equation \ref{eqfrr}) and FAR (defined in Equation \ref{eqfar}), respectively. We also report Half Total Error Rate (HTER), defined in Equation \ref{eqhter}, recommended by Bengio et al. \cite{bengio2002confidence} so we can compare different implementations and attack environments. It is worth noting that under attack environments, the FRRs remain unaffected. Therefore, we report the FRR separately in addition to reporting the FARs and HTERs for each of the adversarial scenarios. 
\begin{equation}
\label{eqfar}
    \textit{FAR} = \frac{\textit{number of successful impostor attempts}}{\textit{total number of impostor attempts}} 
\end{equation}

\begin{equation}
\label{eqfrr}
    \textit{FRR} = \frac{\textit{number of failed genuine attempts}}{\textit{total number of genuine attempts}} 
\end{equation}

\begin{equation}
\label{eqhter}
    \textit{HTER} = \textit{(FAR + FRR)}/2 
\end{equation}

\begin{algorithm} 
\small
\SetAlgoLined
\textbf{Input:} {A[]: list of authentication models for user $u_i \in  U$} \\
\quad    L: length of feature vector to be generated \\
\quad    N: number of feature vectors to be generated \\
\textbf{Output:} {C: A dictionary with keys (authentication models) and values (predicted labels)} \\
C $\gets$ \{ \}, {X'} $\gets$ {[]} \\
 \For{$i \gets 0$ to $N$}{
 {attack\_vector $\gets$ []}\\
 \For{$j  \gets 0$ to $L$}{
    $r \gets \mathcal{U}(0, 1)$\\
    {attack\_vector.append(r)}
 }
 {X'.append(attack\_vector)}
 }
\For{model in $A$}{
{pred\textunderscore labels $\gets$ []} \\
\For{{feat\textunderscore vector in  X'}}{
{pred\textunderscore labels.append(model.predict(feat\textunderscore vector))}}
{C[model] $\gets$ pred\textunderscore labels} \\
}
return C
\caption{{random\textunderscore vector\textunderscore attack}(A[], L, N)}
\label{algo_random_attack}
\end{algorithm} 

\section{Results and Discussion}
\label{ResultsAndDiscussion} 
\begin{figure}[htp]
\centering
\begin{tabular}{c}
\subfigure{\epsfig{file=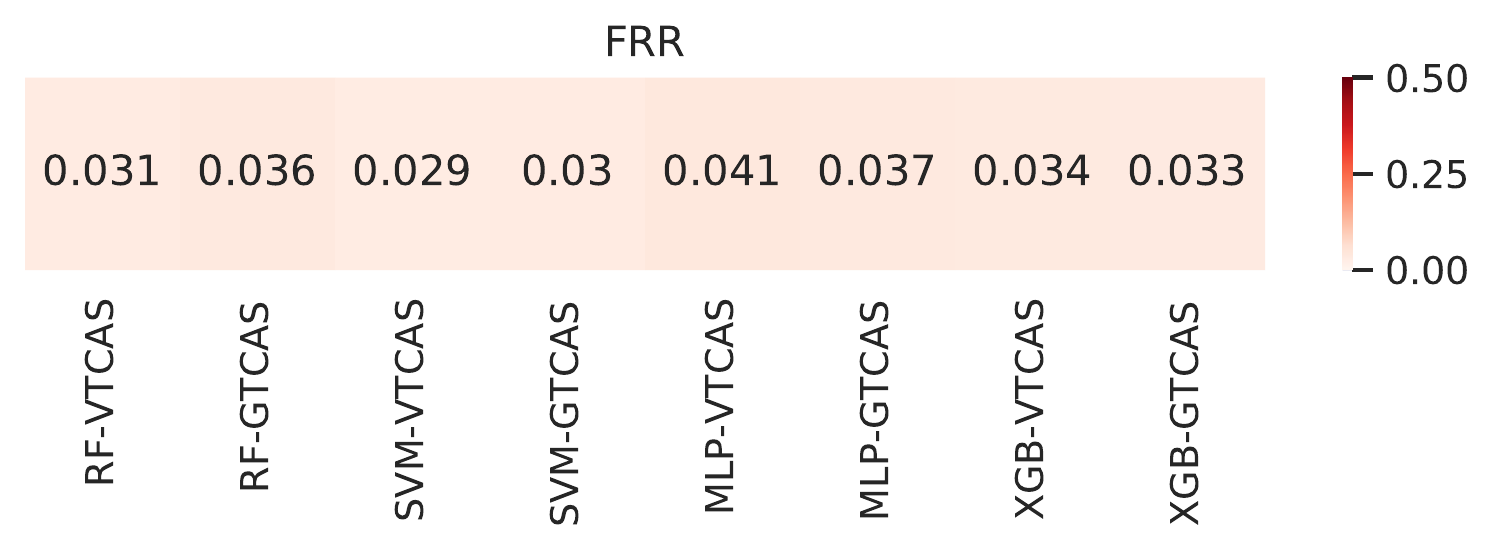, width=3.3in, height=1.15in} \label{BBMAS-FRR}}
\end{tabular}
\caption{FRRs of V-TCAS and G-TCAS for different classifiers.} 
\label{BBMASFRR}
\end{figure}

\begin{figure*}[htp]
\centering
\begin{tabular}{cc}
\subfigure{\epsfig{file=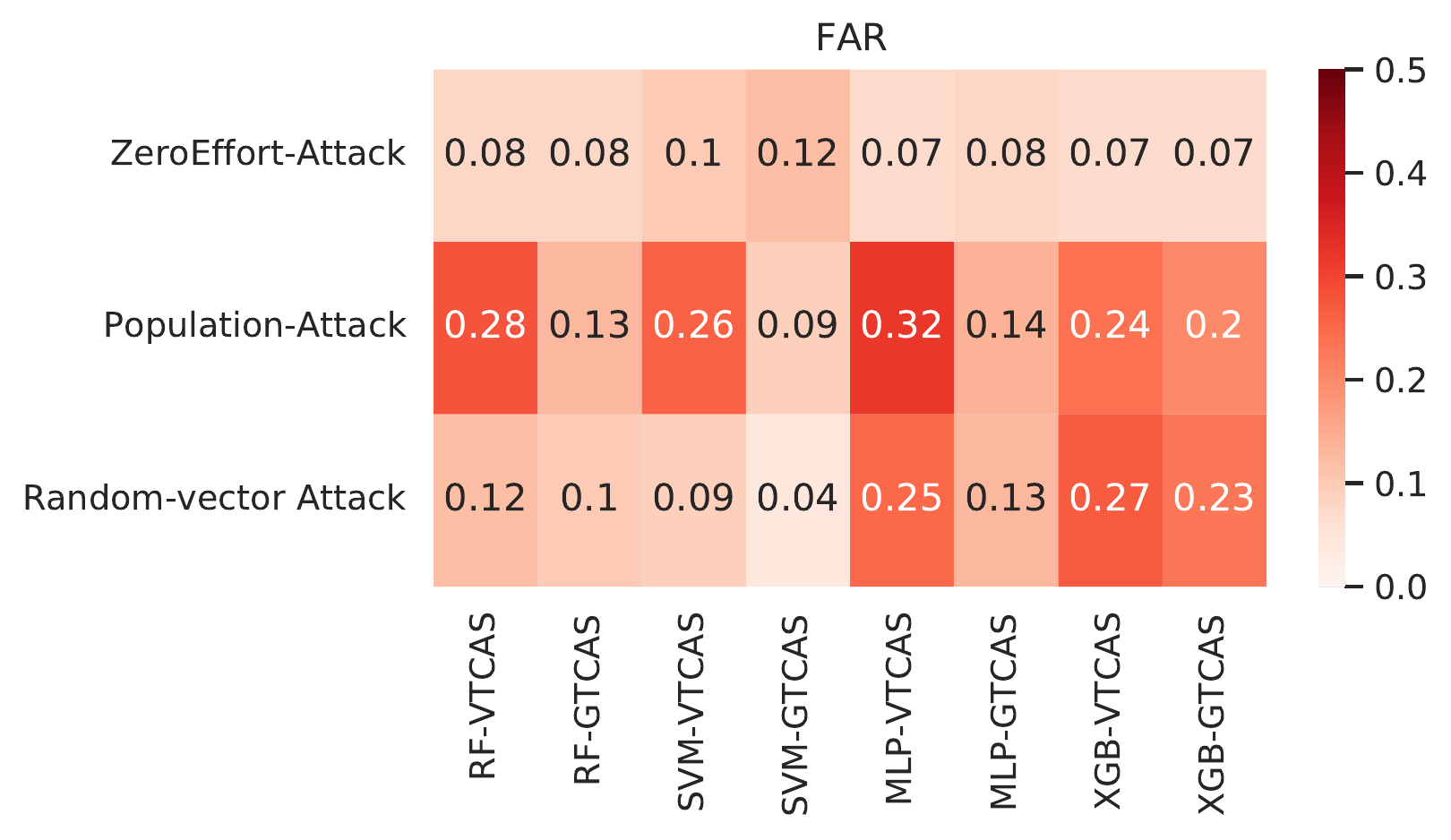, width=3.4in, height=1.7in} \label{BBMAS-FAR1}}
\subfigure{\epsfig{file=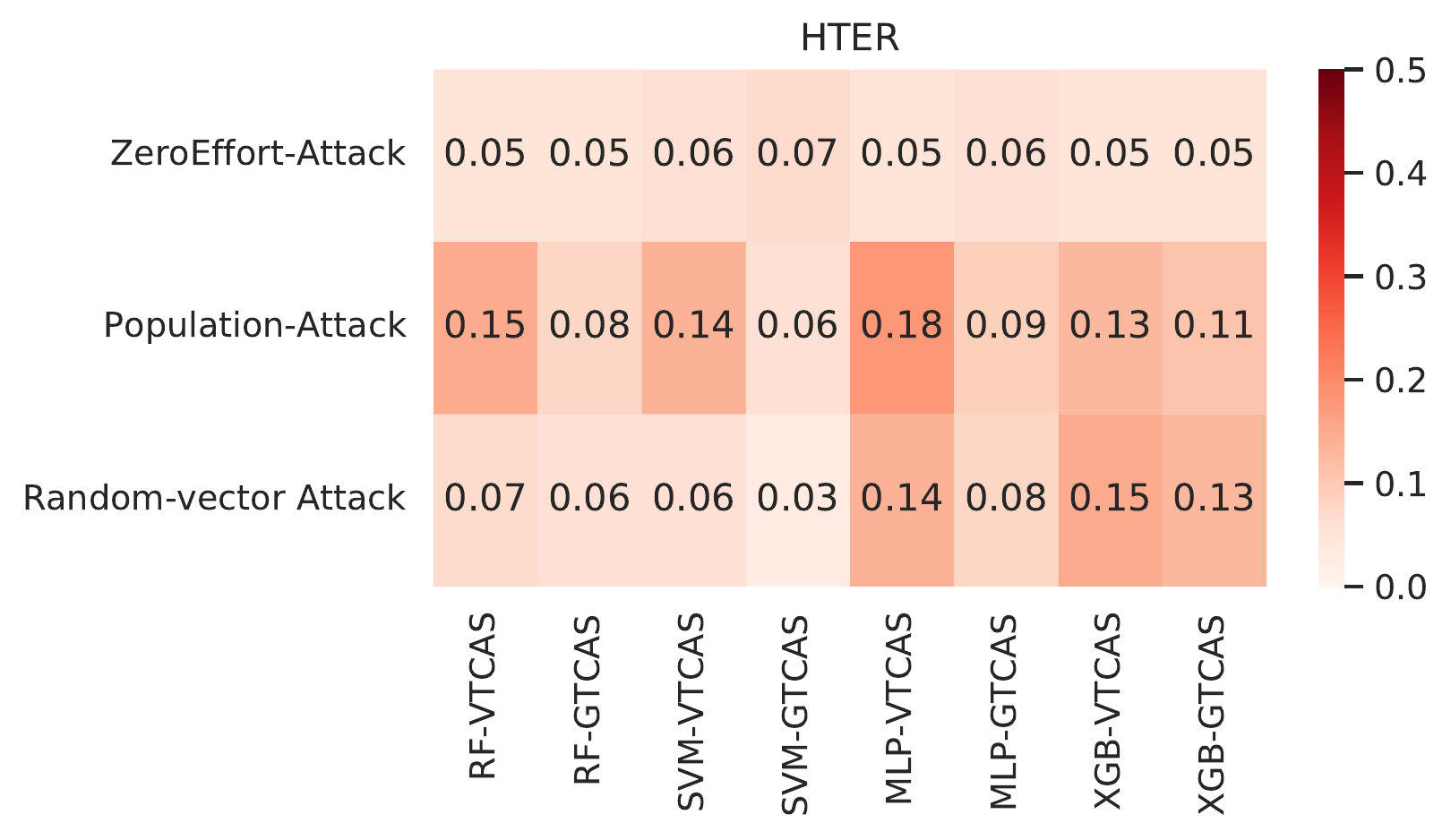, width=3.4in, height=1.7in} \label{BBMAS-HTER1}}
\end{tabular}
\caption{The performance of V-TCAS and G-TCAS under different adversarial scenarios. The benefit of including Genuine and Impostor-GANs in the pipeline, i.e., G-TCAS architecture, is evident as it consistently achieves lower FARs (in turn HTERs) across the classification algorithms and attack scenarios compared to V-TCAS (the architecture without the two GANs).} 
\label{AttackScenarioWise}
\end{figure*}

\begin{figure*}[htp]
\centering
\begin{tabular}{cc}
\subfigure{\epsfig{file=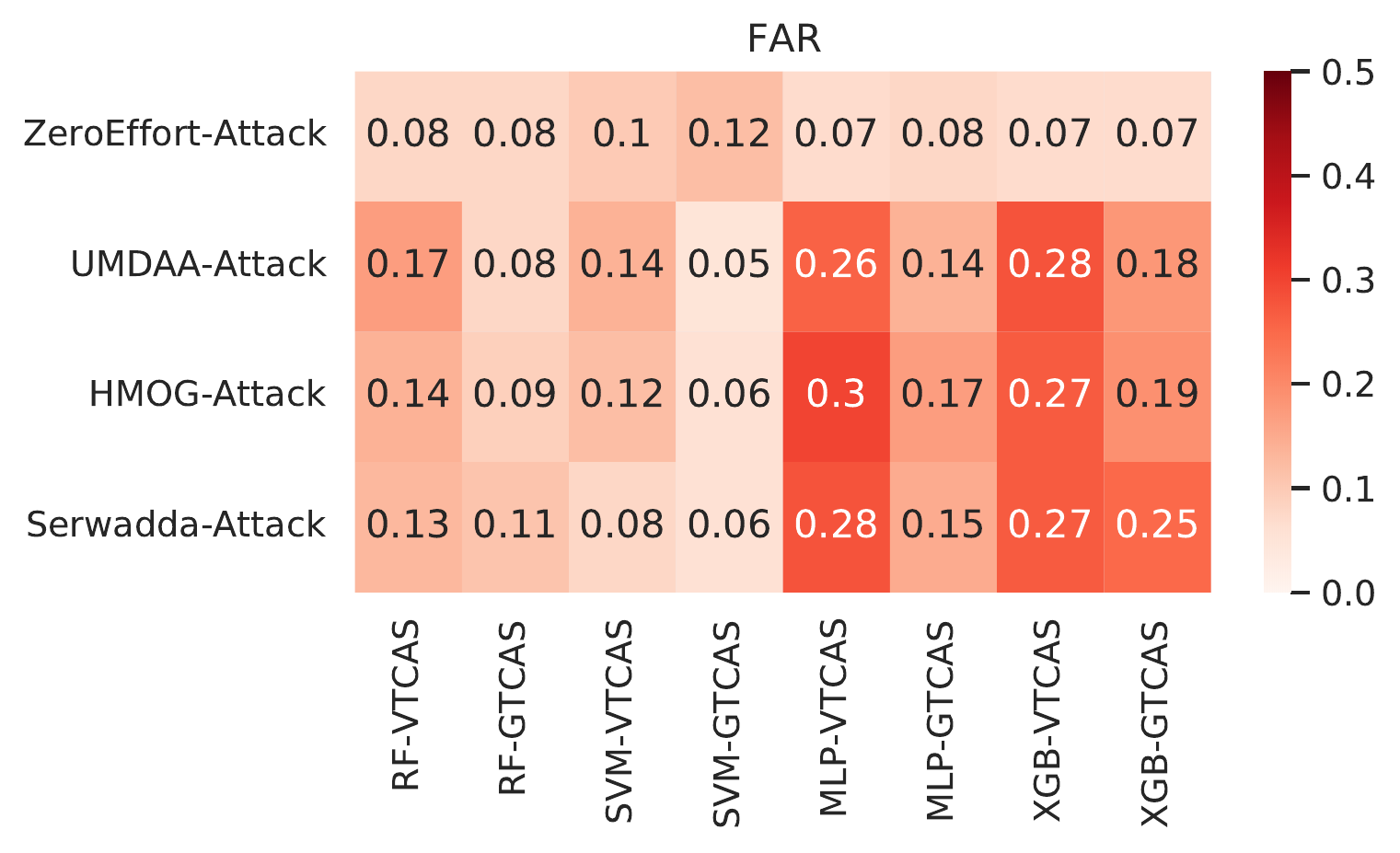, width=3.4in, height=1.7in} \label{BBMAS-FAR2}}
\subfigure{\epsfig{file=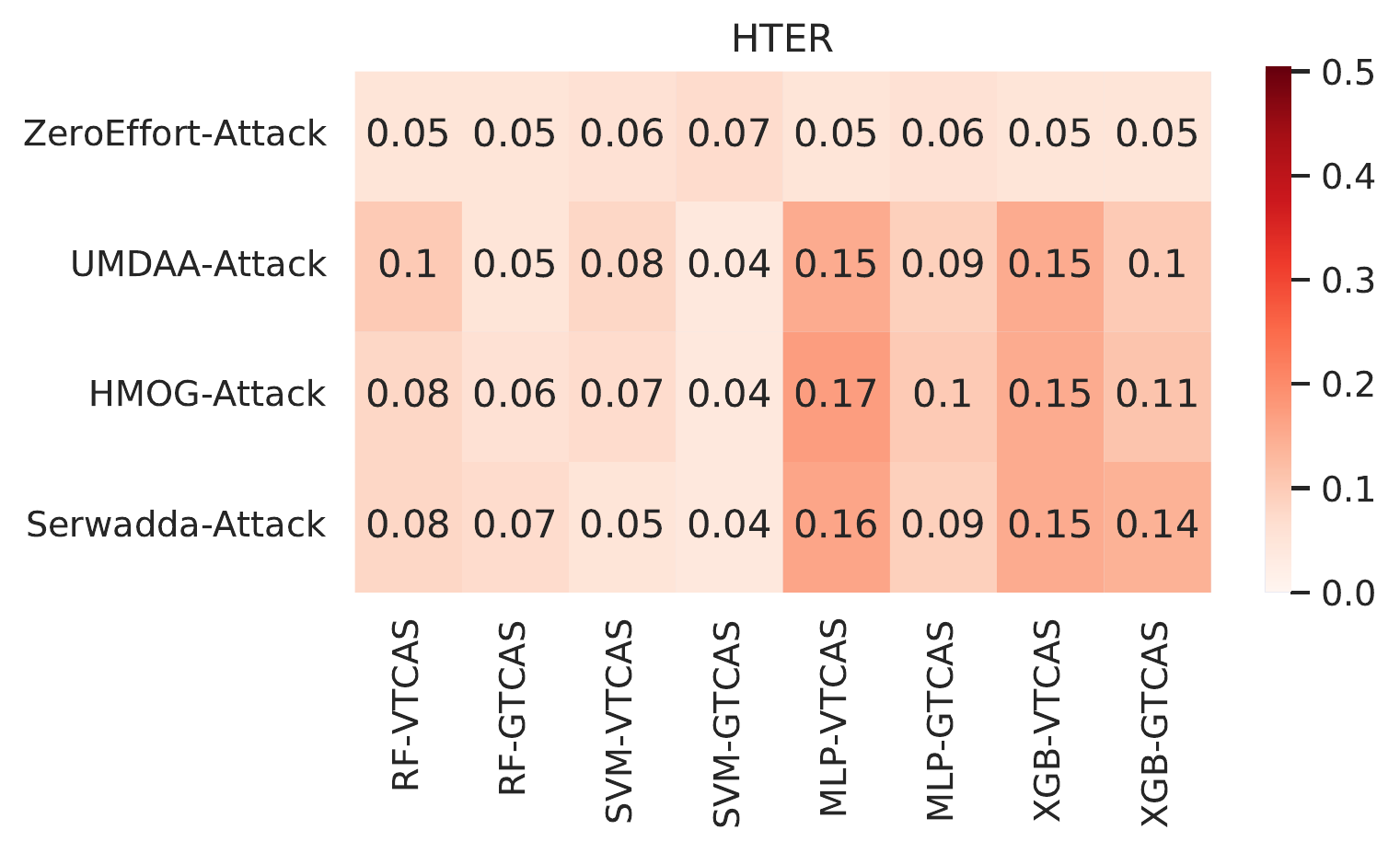, width=3.4in, height=1.7in} 
\label{BBMAS-HTER2}}
\end{tabular}
\caption{The performance of V-TCAS and G-TCAS models under Zero-effort (same dataset) and Zero-effort (cross-dataset) attack scenarios. Interestingly, the Zero-effort attack using a different dataset was more damaging than the one launched using the same dataset. This finding suggests that even for Zero-effort attack scenarios, it is essential that we use different datasets because, in practice, it is very much possible. } 
\label{ZeroEffortSameDifferent}
\end{figure*}

The performance of the TCAS under different adversarial environments is presented in terms of FRR, FAR, and HTER. The FRRs for each of the authentication models remain unaffected across the adversarial scenarios and turned out to be between $2.9-4.5\%$ for different classifiers and architectures (V-TCAS and G-TCAS) (see Figure \ref{BBMASFRR}). FARs and HTERs are presented in Figures \ref{AttackScenarioWise} and \ref{ZeroEffortSameDifferent} across different experimental setups. We break down and present our results for V-TCAS and G-TCAS based on adversarial setups. First, we present the performance of V-TCAS and G-TCAS under Zero-effort (same dataset), Population, and Random-vector attack setups. Further, we present the performance of Zero-effort for same- and cross-dataset attacks. In the end, we present the gender-level analysis of the performance of V-TCAS and G-TCAS.


\begin{figure*}[htp]
\centering
\begin{tabular}{cc}
\subfigure[Zero-effort (same dataset)]{\epsfig{file=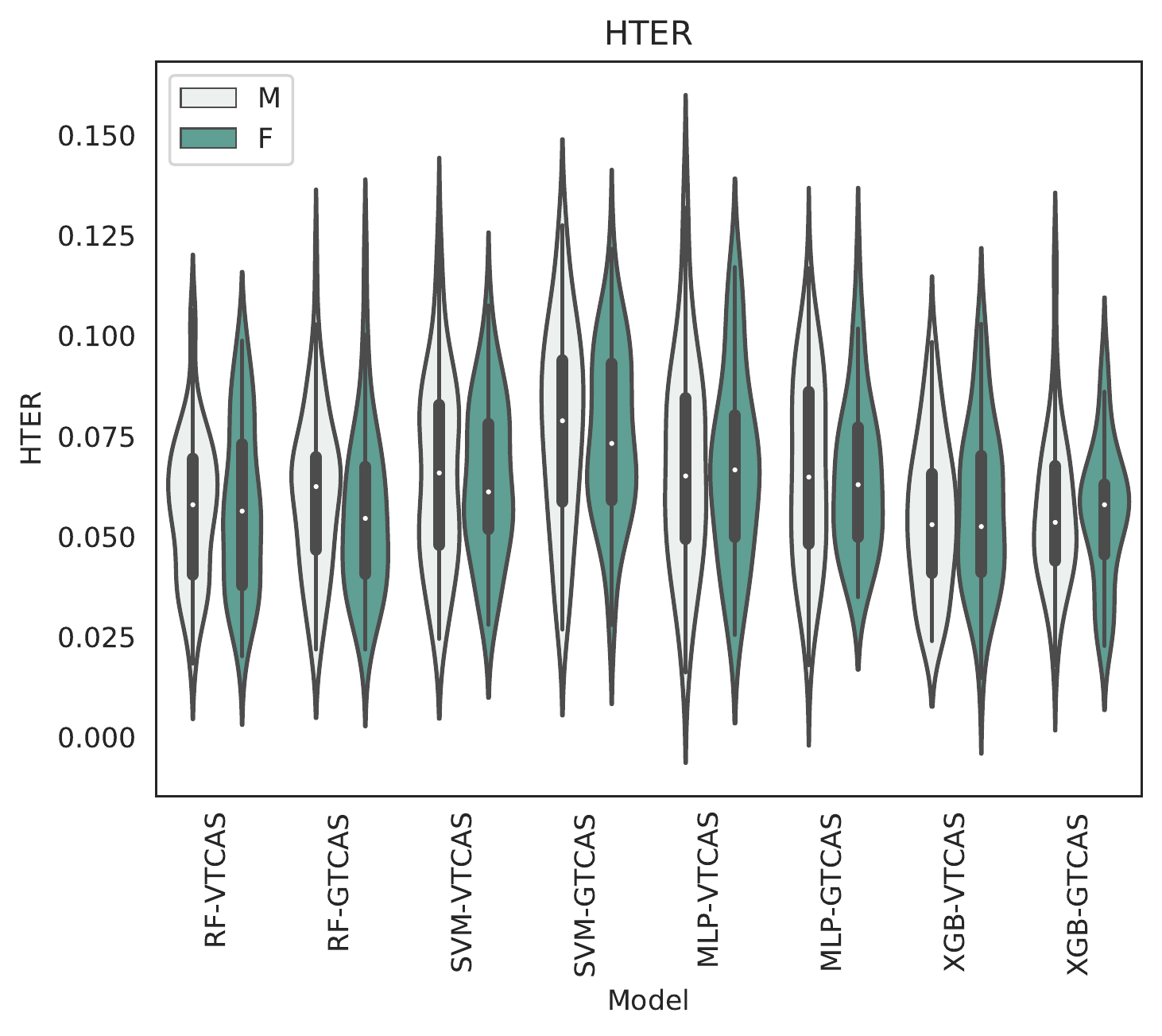, width=3.4in, height=1.8in}
\label{LogReg}} 
\subfigure[Zero-effort (cross dataset average)]{\epsfig{file=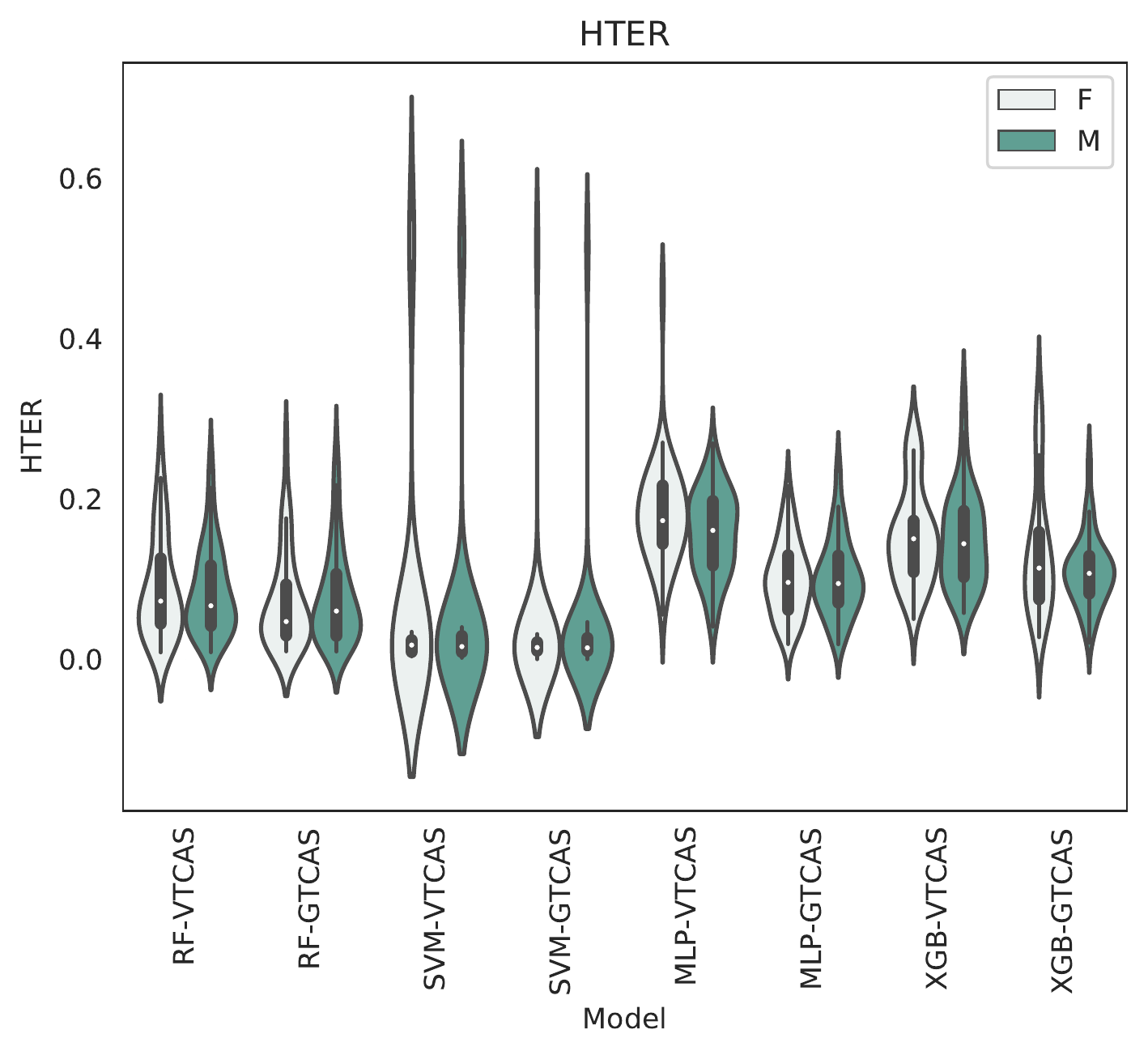, width=3.4in, height=1.8in}
\label{NNet}}\\
\subfigure[Population]{\epsfig{file=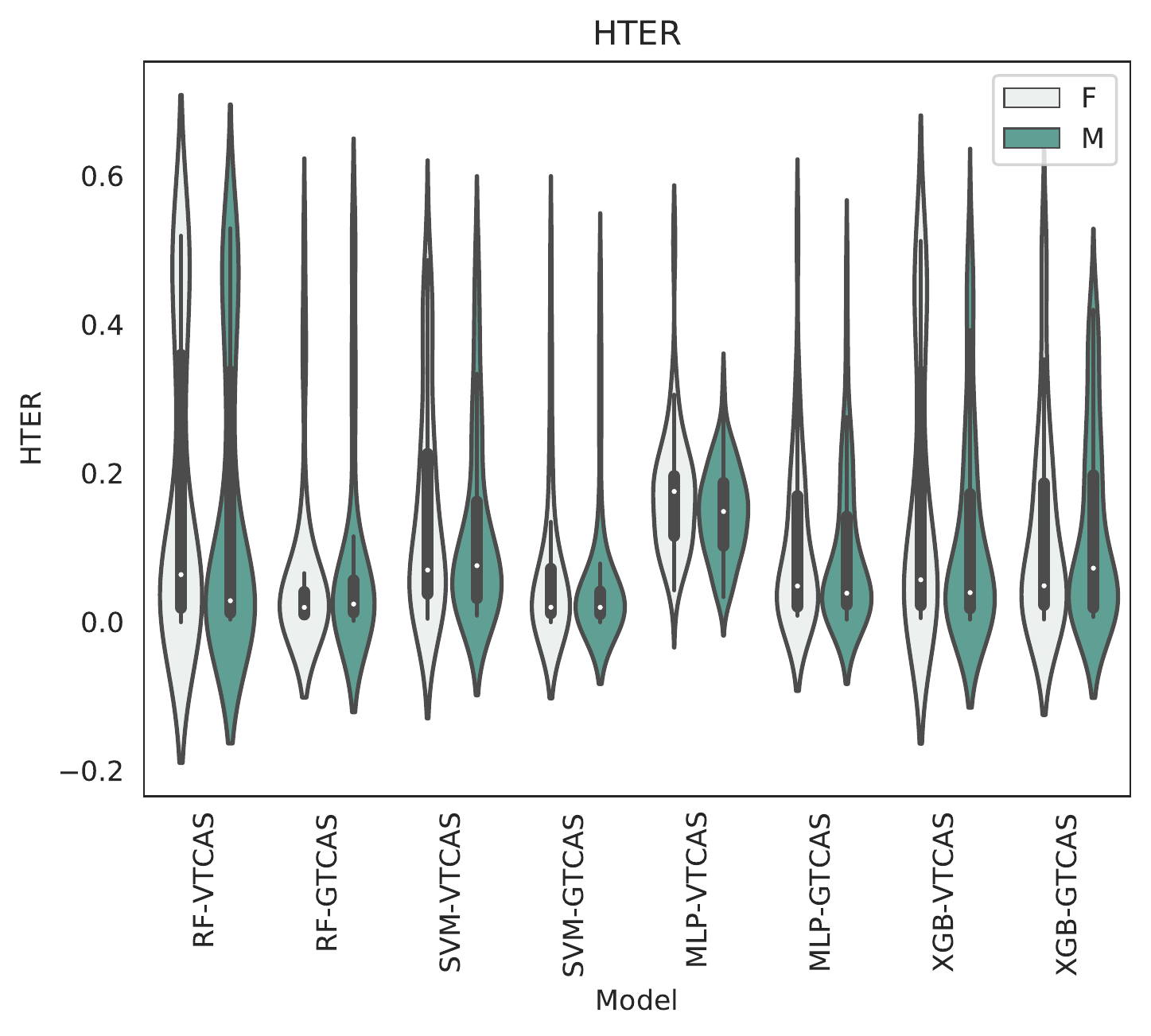, width=3.4in, height=1.8in}
\label{kNN}} 
\subfigure[Random]{\epsfig{file=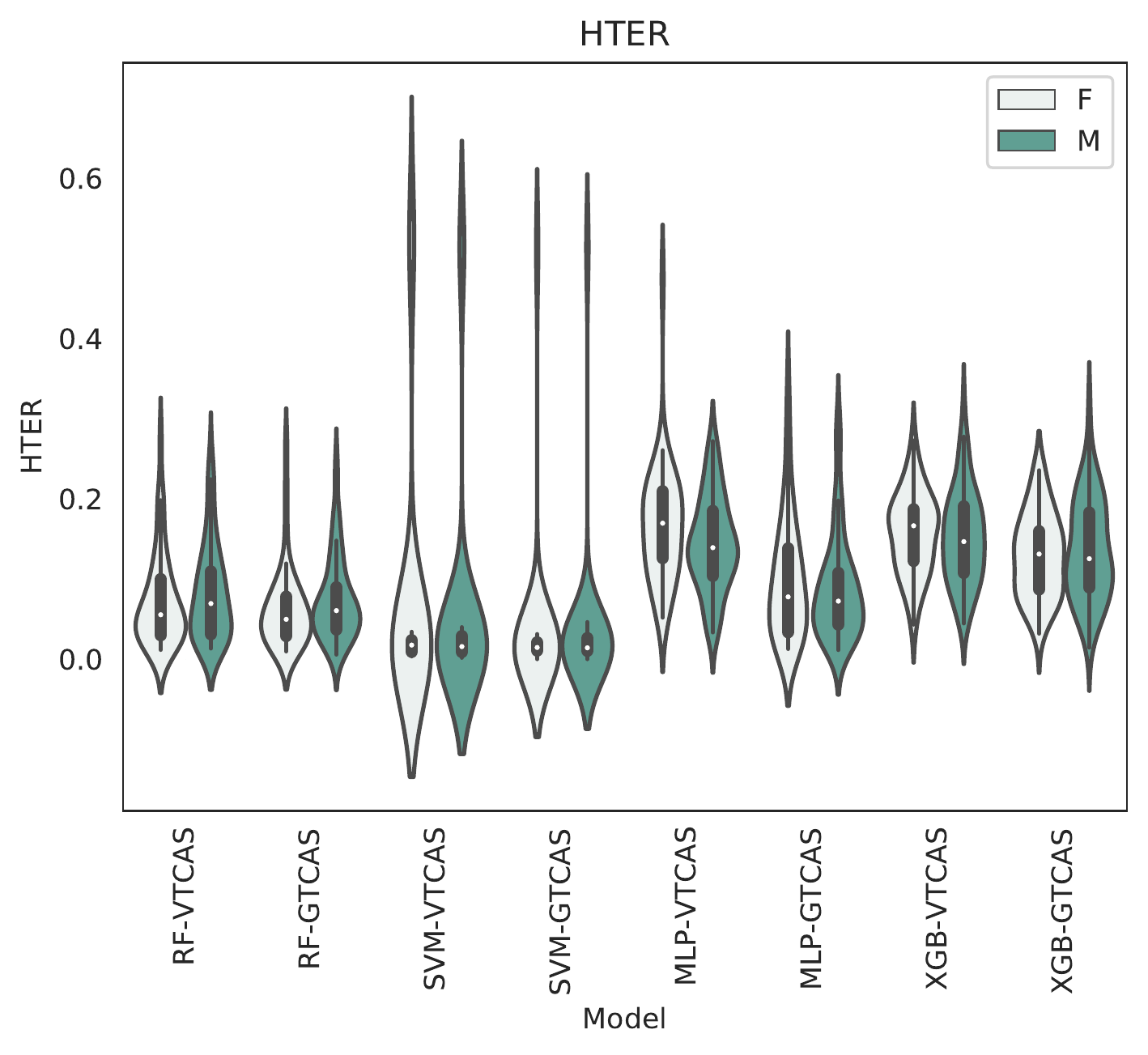, width=3.4in, height=1.8in}
\label{SVM}}
\end{tabular}
\caption{Fairness analysis of V-TCAS and G-TCAS. Its evident from the plots that both architectures (V-TCAS and G-TCAS) regardless of the classification algorithms or adversarial environments achieve similar error rates for across genders. It worth noting that the HTERs seems breaching the boundary of zero primarily because the these plots are kernel density estimation (with Gaussian kernel) of the error rates.}
\label{FairnessAnalysis}
\end{figure*}
\subsection{Zero-effort vs. Population vs. Random-vector} Figure \ref{AttackScenarioWise} summarizes the results obtained under each of the attack scenarios for all classifiers and architectures (V-TCAS and G-TCAS). The TCAS models achieved between $5-7\%$ HTERs and $7-12\%$ FARs for Zero-effort attacks. Population and Random-vector attack severely impacted most V-TCAS models as the FARs increased to $24-34\%$ and $9-27\%$, respectively. In contrast, the G-TCAS models show more resilience than V-TCAS across adversarial scenarios obtaining FARs between $9-20\%$ and $4-23\%$ for Population- and Random-vector attacks, respectively. The FAR heatmap (Figure \ref{BBMAS-FAR1}) suggests that the SVM-G-TCAS is the most robust TCAS architecture, closely followed by RF-G-TCAS, on average. Interestingly, XGB-based models did exceptionally well under the traditional (i.e., the Zero-effort setup) but did not show much resilience under the Population and Random-vector attack scenarios. 

Although the Zero-effort attack serves as an appropriate baseline for comparison with literature \cite{serwadda2013verifiers, frank2012touchalytics} one should not judge the quality of authentication models only based on the performance under the Zero-effort attack scenario. Another inference we can draw from Figure \ref{AttackScenarioWise} is that a Population-based attack is more damaging than the Random-vector and Zero-effort (same dataset) attacks. From this set of results, we encourage future researchers to test their authentication models at least under these three different attack scenarios because the performance evaluated only under the Zero-effort attack scenario could be misleading. 

\subsection{Zero-effort (same) vs. Zero-effort (cross)}
Figure \ref{ZeroEffortSameDifferent} presents the FARs and HTERs under the same dataset Zero-effort attack and cross dataset Zero-effort attacks. As we can see, the cross dataset Zero-effort attacks cause significantly more damage than the same dataset Zero-effort attacks across the classifiers. The heatmap suggests that the traditional, i.e., same dataset, zero-effort attack setup alone is insufficient to evaluate the impostor pass rate for TCAS. Therefore, we recommend that future studies on TCAS evaluate the system under cross dataset Zero-effort attack setup and the same dataset Zero-effort attack setup. 

\begin{figure}[htp]
\centering
\begin{tabular}{c}
\subfigure{\epsfig{file=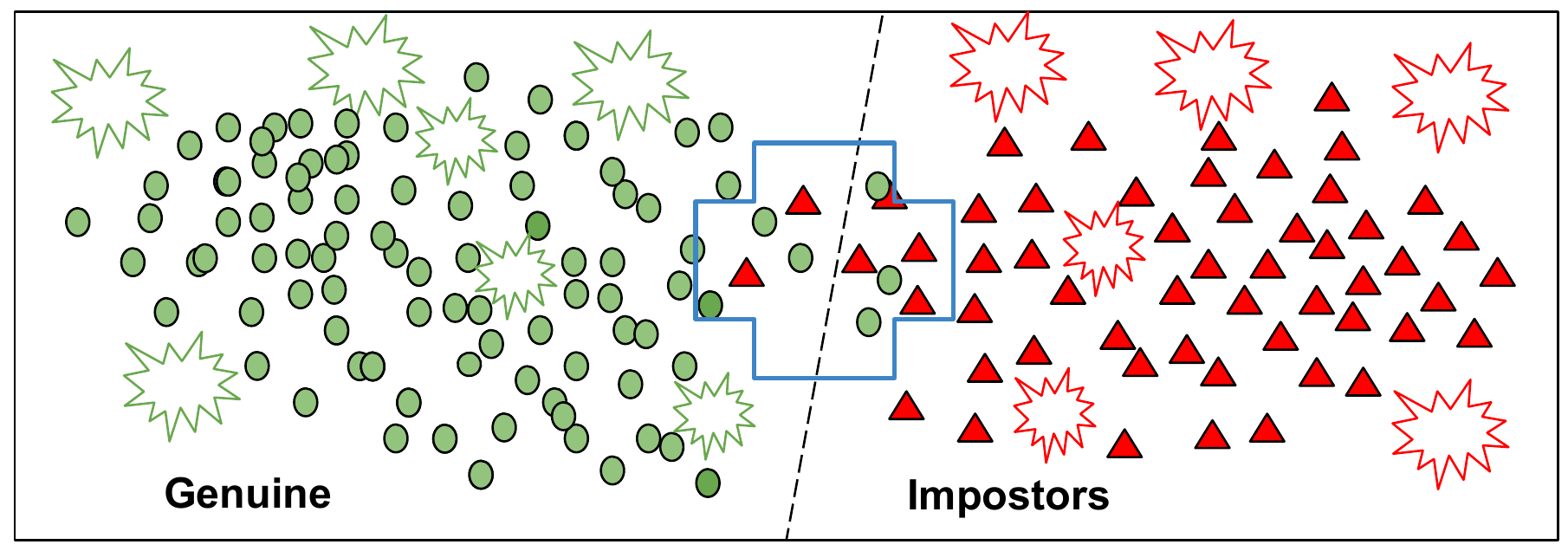, width=3.0in, height=1.0in} \label{NDSS}}
\end{tabular}
\caption{An example of feature space separation by a linear boundary between two classes. An oversimplified version of TCAS demonstrating the acceptance and rejection regions and the overlap area used to compute the error rates. The GANs helped generate more cohesive genuine and impostor data points.} 
\label{ExplainationWhyItWorks}
\end{figure}

More importantly, we can observe that G-TCAS is outstandingly resilient to the cross-dataset Zero-effort attacks compared to V-TCAS consistently across the classifiers. The heatmap clearly shows the importance of including the pair of GANs in the pipeline. In other words, the effectiveness of the proposed architecture G-TCAS is evident. The HTER heatmap suggests that SVM-based G-TCAS is the best architecture, closely followed by Random Forest-based G-TCAS. In comparison, the HTERs achieved by XGB-G-TCAS exceed 10\%, followed by MLP-G-TCAS.  

When multiple datasets are unavailable, the proper way to evaluate the impact of the cross-dataset zero-effort attack is by collecting more data in various experimental settings, from a different user population, or both. Alternatively, one could divide the dataset into multiple groups of users and follow a strategy similar to the k-fold cross-validation. Besides, one can generate multiple datasets algorithmically using population statistics extracted from existing datasets or by altering the distribution of feature values for each user as recommended by Ballard et al. \cite{AlgoAttacksAreMorePractical}.

\subsection{Male vs. Female} One vital aspect that has not been covered much in the TCAS literature is whether TCAS is fair among genders. One of the reasons fairness analysis has not received attention is that most of the public datasets do not contain the gender information of the participants. A recently published dataset, BBMAS-Touch, contained gender information. Therefore, we could conduct a fairness analysis. The results are presented via Figure \ref{FairnessAnalysis}. The error rates across the TCAS architectures and adversarial scenarios suggest that TCAS does not discriminate among different genders. This conclusion, however, is limited by a limited user dataset. The fairness aspect of TCAS needs to be paid attention to and studied further for people of different demographics. In the future, we would like to include a recently published dataset such as \cite{Acien2021} which consists of demographic information of $600$ users in the fairness analysis.

\subsection{Motivation Behind the G-TCAS Architecture}
In the conference paper \cite{IJCB2021}, we demonstrated that the inclusion of synthetic genuine and impostor data generated by Genuine- and Impostor-GANs helped separate the data better. Consequently, G-TCAS showed more resilience than V-TCAS. Further discussion on the motivation is drawn from \cite{RandomAttackMutibiometric}. Technically, TCAS is trained to classify a particular region (aka acceptance region) as genuine and a separate region as an impostor (aka rejection region). 
Although Figure \ref{ExplainationWhyItWorks} presents an oversimplified scenario, it provides insight into the success of the adversarial scenarios besides the Zero-effort (same) included in this paper. As depicted in Figure \ref{ExplainationWhyItWorks}, the Genuine-GAN helped us fill in the green stars, and Impostor-GANs helped us fill in the red stars. The classifiers, thus, were able to draw a better boundary. Recently published datasets such as \cite{Acien2021} consisting of a significantly high number of users, gestures per user, and type of devices used in the experiment, can be used to train the generative methods better.

\section{Conclusion and Future Work}
\label{secConclusionAndFutureWork}
We evaluated V-TCAS and G-TCAS under three active adversarial environments. G-TCAS showed significantly more resilience than V-TCAS across the classifiers and adversarial environments. We also found that traditionally studied Zero-effort attack does more damage if launched using a different dataset than the same dataset the genuine user comes from. In addition, we found that TCAS is not unfair to different genders. We evaluated V-TCAS and G-TCAS under a variety of minimal effort attacks. In the future, we will assess the robustness of V-TCAS and G-TCAS under moderate and high-effort attacks. In addition, we aim to investigate whether the idea of including dual GAN in the classification pipeline extends to other behavioral biometrics such as sensor-based gait. Further, we would like to study the effectiveness of variations of GAN such as Composite Travel Generative Adversarial Networks (CT-GAN) and AC-GAN. Moreover, we plan to use explain-ability tools to open the pipeline and show why G-TCAS is more resilient than V-TCAS. 




\ifCLASSOPTIONcompsoc
  \section*{Acknowledgments}
\else
  \section*{Acknowledgment}
\fi
We would like to thank anonymous reviewers for providing invaluable feedback. Rajiv Ratn Shah was partly supported by the Infosys Center for Artificial Intelligence and the Center of Design and New Media at IIIT Delhi, India.

\bibliography{short_references.bib}

\begin{thebibliography}{10}

\bibitem{OneClassRajesh2018}
Rajesh Kumar, Partha~P. Kundu, and Vir~V. Phoha.
\newblock Continuous authentication using one-class classifiers and their
  fusion.
\newblock In {\em IEEE ISBA}, 2018.

\bibitem{kumar2016continuous}
Rajesh Kumar, Vir~V Phoha, and Abdul Serwadda.
\newblock Continuous authentication of smartphone users by fusing typing,
  swiping, and phone movement patterns.
\newblock In {\em IEEE BTAS}, 2016.

\bibitem{VoiceContAuth}
Huan Feng, Kassem Fawaz, and Kang~G Shin.
\newblock Continuous authentication for voice assistants.
\newblock In {\em MobiCom}, 2017.

\bibitem{IJCB2017}
Rajesh Kumar, Partha~Pratim Kundu, Diksha Shukla, and Vir~V Phoha.
\newblock Continuous user authentication via unlabeled phone movement patterns.
\newblock In {\em IEEE IJCB}, 2017.

\bibitem{frank2012touchalytics}
Mario Frank, Ralf Biedert, Eugene Ma, Ivan Martinovic, and Dawn Song.
\newblock Touchalytics: On the applicability of touchscreen input as a
  behavioral biometric for continuous authentication.
\newblock {\em IEEE T-IFS}, 2012.

\bibitem{TouchFirstAuth}
Tao Feng, Ziyi Liu, Kyeong-An Kwon, Weidong Shi, Bogdan Carbunar, Yifei Jiang,
  and Nhung Nguyen.
\newblock Continuous mobile authentication using touchscreen gestures.
\newblock In {\em IEEE HST}, 2012.

\bibitem{serwadda2013verifiers}
Abdul Serwadda, Vir~V Phoha, and Zibo Wang.
\newblock Which verifiers work?: A benchmark evaluation of touch-based
  authentication algorithms.
\newblock In {\em IEEE BTAS}, 2013.

\bibitem{ModalSwipeContinuous}
Soumik Mondal and Patrick Bours.
\newblock Swipe gesture based continuous authentication for mobile devices.
\newblock In {\em IEEE ICB}, 2015.

\bibitem{patel2016continuous}
Vishal~M Patel, Rama Chellappa, Deepak Chandra, and Brandon Barbello.
\newblock Continuous user authentication on mobile devices: Recent progress and
  remaining challenges.
\newblock {\em IEEE Signal Processing Magazine}, 2016.

\bibitem{sitova2015hmog}
Zde{\v{n}}ka Sitov{\'a}, Jaroslav {\v{S}}ed{\v{e}}nka, Qing Yang, Ge~Peng, Gang
  Zhou, Paolo Gasti, and Kiran~S Balagani.
\newblock Hmog: New behavioral biometric features for continuous authentication
  of smartphone users.
\newblock {\em IEEE T-IFS}, 2015.

\bibitem{lee2017method}
Ruby~B Lee and Wei-Han Lee.
\newblock Method and system for implicit authentication, August~10 2017.
\newblock US Patent App. 15/428,306.

\bibitem{lee2017methodpaper}
Wei-Han Lee and Ruby~B. Lee.
\newblock Implicit smartphone user authentication with sensors and contextual
  machine learning.
\newblock In {\em 2017 47th Annual IEEE/IFIP-DSN}, 2017.

\bibitem{2018Benchmark}
Julian Fierrez, Ada Pozo, Marcos Martinez-Diaz, Javier Galbally, and Aythami
  Morales.
\newblock Benchmarking touchscreen biometrics for mobile authentication.
\newblock {\em IEEE T-IFS}, 2018.

\bibitem{BehavioralBiometricsAcceptability}
Andreas Skalkos, Ioannis Stylios, Maria Karyda, and Spyros Kokolakis.
\newblock Users’ privacy attitudes towards the use of behavioral biometrics
  continuous authentication (bbca) technologies: A protection motivation theory
  approach.
\newblock {\em Journal of Cybersecurity and Privacy}, 2021.

\bibitem{ForgeryResistantTCASFrank}
Neil~Zhenqiang Gong, Mathias Payer, Reza Moazzezi, and Mario Frank.
\newblock Forgery-resistant touch-based authentication on mobile devices.
\newblock In {\em AsiaCCS}, 2016.

\bibitem{Li2013UnobservableRF}
Lingjun Li, Xinxin Zhao, and Guoliang Xue.
\newblock Unobservable re-authentication for smartphones.
\newblock In {\em NDSS}, 2013.

\bibitem{RamaChellappa}
Upal Mahbub, Sayantan Sarkar, Vishal~M Patel, and Rama Chellappa.
\newblock Active user authentication for smartphones: A challenge data set and
  benchmark results.
\newblock In {\em IEEE BTAS}, 2016.

\bibitem{bengio2002confidence}
Samy Bengio, Christine Marcel, Sebastien Marcel, and Johnny Mari{\'e}thoz.
\newblock Confidence measures for multimodal identity verification.
\newblock {\em Information Fusion}, 2002.

\bibitem{poh2006HTER}
Norman Poh and Samy Bengio.
\newblock Database, protocols and tools for evaluating score-level fusion
  algorithms in biometric authentication.
\newblock {\em Pattern Recognition}, 2006.

\bibitem{chellappa2019continuous}
Upal Mahbub, Jukka Komulainen, Denzil Ferreira, and Rama Chellappa.
\newblock Continuous authentication of smartphones based on application usage.
\newblock {\em IEEE T-BIOM}, 2019.

\bibitem{EffortAsAFactor}
NIST.
\newblock Presentation attack detection (pad).
\newblock \url{https://pages.nist.gov/SOFA/SOFA.html}, 2022.
\newblock Online; accessed February 28, 2022.

\bibitem{RandomAttackMutibiometric}
Benjamin Zi~Hao Zhao, Hassan~Jameel Asghar, and Mohamed~Ali Kaafar.
\newblock On the resilience of biometric authentication systems against random
  inputs.
\newblock {\em NDSS}, 2020.

\bibitem{MimicryAttackOnSwipes}
Hassan Khan, Urs Hengartner, and Daniel Vogel.
\newblock Targeted mimicry attacks on touch input based implicit authentication
  schemes.
\newblock In {\em ACM MobiSys}, 2016.

\bibitem{RoboticRobbery}
Abdul Serwadda, Vir~V Phoha, Zibo Wang, Rajesh Kumar, and Diksha Shukla.
\newblock Toward robotic robbery on the touch screen.
\newblock {\em ACM TISSEC (now TOPS)}, 2016.

\bibitem{AttackSurvey2021}
Ren{\'e} Mayrhofer and Stephan Sigg.
\newblock Adversary models for mobile device authentication.
\newblock {\em ACM Computing Survey}, 2021.

\bibitem{populationattackGait}
Tiantian Zhu, Lei Fu, Qiang Liu, Zi~Lin, Yan Chen, and Tieming Chen.
\newblock One cycle attack: Fool sensor-based personal gait authentication with
  clustering.
\newblock {\em IEEE T-IFS}, 2020.

\bibitem{Snoop-forge-reply-keystroke}
Khandaker~A Rahman, Kiran~S Balagani, and Vir~V Phoha.
\newblock Snoop-forge-replay attacks on continuous verification with
  keystrokes.
\newblock {\em IEEE T-IFS}, 2013.

\bibitem{DistanceBasedMatchersAreImmuneToRandomAttacks}
Elena Pagnin, Christos Dimitrakakis, Aysajan Abidin, and Aikaterini Mitrokotsa.
\newblock On the leakage of information in biometric authentication.
\newblock In {\em Indocrypt}. Springer, 2014.

\bibitem{rajesh2020}
Rajesh Kumar, Can Isik, and Vir~V Phoha.
\newblock Treadmill assisted gait spoofing (tags): An emerging threat to
  wearable sensor-based gait authentication.
\newblock {\em ACM DTRAP}, 2021.

\bibitem{MyPhDThesis}
Rajesh Kumar.
\newblock Treadmill assisted circumvention of wearable sensors-based gait
  authentication, phd thesis.
\newblock {\em Syracuse University, USA}, 2021.

\bibitem{LatestSurveyTouchUsabilitySecurity2020}
Elakkiya Ellavarason, Richard Guest, Farzin Deravi, Raul Sanchez-Riello, and
  Barbara Corsetti.
\newblock Touch-dynamics based behavioural biometrics on mobile devices--a
  review from a usability and performance perspective.
\newblock {\em ACM Computing Surveys}, 2020.

\bibitem{goodfellow2014generative}
Ian Goodfellow, Jean Pouget-Abadie, Mehdi Mirza, Bing Xu, David Warde-Farley,
  Sherjil Ozair, Aaron Courville, and Yoshua Bengio.
\newblock Generative adversarial nets.
\newblock {\em NeurIPS}, 2014.

\bibitem{IJCB2021}
Mohit Agrawal, Pragyan Mehrotra, Rajesh Kumar, and Rajiv~Ratn Shah.
\newblock Defending touch-based continuous authentication systems from active
  adversaries using generative adversarial networks.
\newblock In {\em IEEE IJCB}, 2021.

\bibitem{GANTouch2020}
Debzani Deb and Mina~M Guirguis.
\newblock Use of auxiliary classifier generative adversarial network in
  touchstroke authentication.
\newblock In {\em ICMLA}, 2020.

\bibitem{GANBA2022}
Alejandro Gomez-Alanis, Jose~A Gonzalez-Lopez, and Antonio~M Peinado.
\newblock Ganba: Generative adversarial network for biometric anti-spoofing.
\newblock {\em Applied Sciences}, 2022.

\bibitem{DefenseGAN2018}
Pouya Samangouei, Maya Kabkab, and Rama Chellappa.
\newblock Defense-gan: Protecting classifiers against adversarial attacks using
  generative models.
\newblock {\em ICLR}, abs/1805.06605, 2018.

\bibitem{belman2019insights}
Amith K~Belman et~al.
\newblock Insights from bb-mas--a large dataset for typing, gait and swipes of
  the same person on desktop, tablet and phone.
\newblock {\em arXiv preprint arXiv:1912.02736}, 2019.

\bibitem{he2008adasyn}
Haibo He, Yang Bai, Edwardo~A Garcia, and Shutao Li.
\newblock Adasyn: Adaptive synthetic sampling approach for imbalanced learning.
\newblock In {\em IEEE IJCNN}, 2008.

\bibitem{chawla2002smote}
Nitesh~V Chawla, Kevin~W Bowyer, Lawrence~O Hall, and W~Philip Kegelmeyer.
\newblock Smote: synthetic minority over-sampling technique.
\newblock {\em JAIR}, 2002.

\bibitem{AlgoAttacksAreMorePractical}
Lucas Ballard, Daniel Lopresti, and Fabian Monrose.
\newblock Forgery quality and its implications for behavioral biometric
  security.
\newblock {\em IEEE T-SMC, Part B (Cybernetics)}, 2007.

\bibitem{Acien2021}
Alejandro Acien, Aythami Morales, Julian Fierrez, Ruben Vera-Rodriguez, and
  Oscar Delgado~Mohatar.
\newblock Becaptcha: Behavioral bot detection using touchscreen and mobile
  sensors benchmarked on humidb.
\newblock {\em Engineering Applications of Artificial Intelligence}, 2021.

\end{thebibliography}
\bibliographystyle{unsrt}
\vspace{0.15in}
\hangindent=\dimexpr 1.2in+1em\relax
\hangafter=-10
\noindent\llap{\raisebox{\dimexpr\ht\strutbox-\height}[0pt][0pt]{\includegraphics[width=1in,height=0.9in]{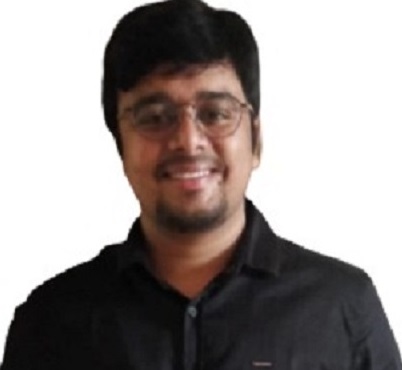}}%
\hspace{1em}}
\scriptsize \textbf{Mohit Agrawal} is a Ph.D. student at IIIT Delhi, India. He also works as a Senior Machine Learning Engineer at Qualcomm. He was associated with India Naval Academy as a guest faculty member. He earned his Master's degree in Computer Science and Engineering from the National Institute of Technology, Tiruchirappalli. His research interest includes Biometrics, Social Network Analysis, Computer Vision, and Machine Learning. \\

\hangindent=\dimexpr 1.1in+1em\relax
\hangafter=-9
\noindent\llap{\raisebox{\dimexpr\ht\strutbox-\height}[0pt][0pt]{\includegraphics[width=1in,height=1.15in]{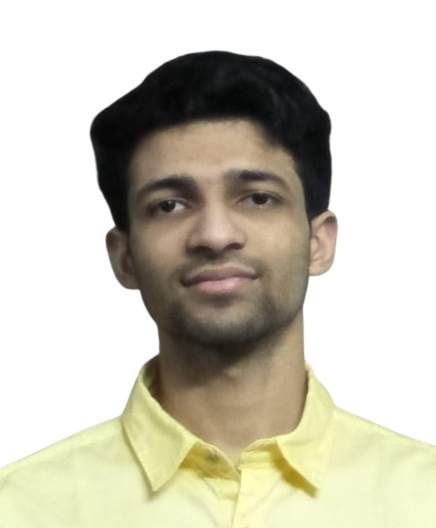}}%
\hspace{1em}}
\scriptsize \textbf{Pragyan Mehrotra} is a software engineer at Amazon Dublin, Ireland. He received his B.Tech in Computer Science and Engineering from the IIIT, Delhi, India. During his bachelor's degree, he has also worked as a software engineering intern at Microsoft Bangalore, India. His area of interest includes Biometrics, Machine Learning, Deep Learning, Cybersecurity, and Image Processing. \\

\hangindent=\dimexpr 1.1in+1em\relax
\hangafter=-10
\noindent\llap{\raisebox{\dimexpr\ht\strutbox-\height}[0pt][0pt]{\includegraphics[width=1in,height=1in]{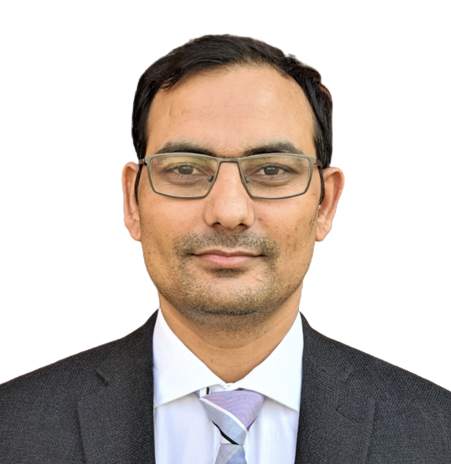}}%
\hspace{1em}}
\scriptsize \textbf{Rajesh Kumar} is an Assistant Professor at Bucknell University, USA. Before joining Bucknell University, he was an Assistant Professor at Hofstra University, USA, and a visiting faculty at Haverford College, USA. He earned his Ph.D. in Computer and Information Science and Engineering (CISE) from Syracuse University, USA, after receiving his masters in Mathematics from Louisiana Tech University, USA, and in Computer Applications from Jawaharlal Nehru University India. His research focuses on harnessing the power of ever-evolving smart devices, wearables, and machine intelligence to solve security, privacy, and healthcare problems. \\

\hangindent=\dimexpr 1.1in+1em\relax
\hangafter=-10
\noindent\llap{\raisebox{\dimexpr\ht\strutbox-\height}[0pt][0pt]{\includegraphics[width=1in,height=1in]{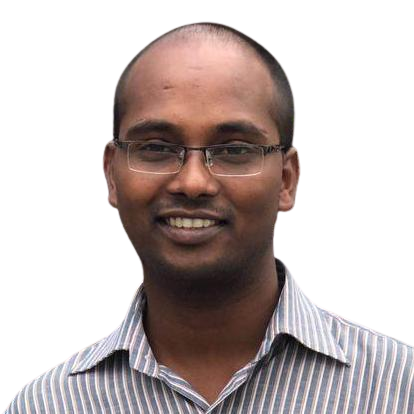}}%
\hspace{1em}}
\scriptsize \textbf{Rajiv Ratn Shah} is an Assistant Professor in the Department of Computer Science and Engineering (joint appointment with the Department of Human-centered Design) at IIIT-Delhi. He is the founder of the MIDAS lab at IIIT-Delhi. He received his Ph.D. in Computer Science from the National University of Singapore, Singapore. Dr. Shah is the recipient of several awards, including the prestigious Heidelberg Laureate Forum (HLF) and European Research Consortium for Informatics and Mathematics (ERCIM) fellowships. His research interests include multimedia content processing, natural language processing, image processing, multimodal computing, data science, social media computing, and the internet of things.
\end{document}